\documentclass[oneside,11pt]{article}

\usepackage[utf8]{inputenc}
\usepackage[T1]{fontenc}

\usepackage{amsthm}
\usepackage{amsmath}
\usepackage{amsfonts}
\usepackage{amssymb}

\usepackage{graphicx}
\usepackage{float}
\usepackage{booktabs}
\usepackage{multirow}
\usepackage{rotating}
\usepackage{array}
\usepackage{xcolor}

\usepackage{breakcites}

\usepackage{enumitem}

\usepackage[top=1.5cm,bottom=2cm,right=2.5cm,left=2.5cm]{geometry}
\usepackage{titling}
\usepackage{natbib}

\usepackage{ifplatform}

\usepackage{textcomp}

\ifwindows
\fi

\ifwindows
	\usepackage{bbm}
\else
	\iflinux
		\usepackage{bbm}
	\else
		\usepackage{bbm}
	\fi
\fi

\newcommand{\lp}{\left(}
\newcommand{\rp}{\right)}

\newcommand{\lb}{\left\{}
\newcommand{\rb}{\right\}}

\newcommand{\R}{\mathbb{R}}
\newcommand{\T}{\mathbb{T}}

\newcommand{\Z}{\mathbb{Z}}

\newcommand{\rd}{\mathrm{d}}

\newcommand{\bx}{\mathbf{x}}

\newcommand{\bm}{\mathbf{m}}
\newcommand{\bn}{\mathbf{n}}
\newcommand{\by}{\mathbf{y}}
\newcommand{\bX}{\mathbf{X}}
\newcommand{\bY}{\mathbf{Y}}
\newcommand{\bZ}{\mathbf{Z}}
\newcommand{\bJ}{\mathbf{J}}
\newcommand{\bU}{\mathbf{U}}
\newcommand{\bF}{\mathbf{F}}
\newcommand{\bmu}{\boldsymbol\mu}
\newcommand{\bu}{\mathbf{u}}
\newcommand{\bv}{\mathbf{v}}
\newcommand{\bb}{\mathbf{b}}

\newcommand{\bs}{\mathbf{s}}
\newcommand{\bz}{\mathbf{z}}
\newcommand{\bk}{\mathbf{k}}
\newcommand{\bp}{\mathbf{p}}

\newcommand{\zero}{\mathbf{0}}

\newcommand{\bphi}{\boldsymbol\varphi}
\newcommand{\ba}{\boldsymbol\alpha}

\newcommand{\bthe}{\boldsymbol\theta}
\newcommand{\btheta}{\boldsymbol\theta}
\newcommand{\bkappa}{\boldsymbol\kappa}
\newcommand{\bTheta}{\boldsymbol\Theta}
\newcommand{\Ical}{\mathcal{I}}
\newcommand{\bsigma}{\boldsymbol\sigma}
\newcommand{\bSigma}{\boldsymbol\Sigma}
\newcommand{\bLambda}{\boldsymbol\Lambda}
\newcommand{\bGamma}{\boldsymbol\Gamma}

\newcommand{\bHcal}{\boldsymbol{\mathcal{H}}}

\newcommand{\blambda}{\boldsymbol\lambda}
\newcommand{\bB}{\mathbf{B}}
\newcommand{\bK}{\mathbf{K}}
\newcommand{\bA}{\mathbf{A}}
\newcommand{\bV}{\mathbf{V}}
\newcommand{\bI}{\mathbf{I}}
\newcommand{\bS}{\mathbf{S}}
\newcommand{\bH}{\mathbf{H}}

\newcommand{\bM}{\mathbf{M}}
\newcommand{\bW}{\mathbf{W}}

\newcommand{\lrp}[1]{\left(#1\right)}
\newcommand{\lrc}[1]{\left[#1\right]}
\newcommand{\lrb}[1]{\left\{#1\right\}}
\newcommand{\Prob}[1]{\mathbb{P}\lb #1\rb}

\newcommand{\diag}[1]{\mathrm{diag}\lp #1\rp}
\newcommand{\cmod}[1]{\mathrm{cmod}\left(#1\right)}

\newcommand{\pf}[2]{\frac{\partial #1}{\partial #2}}
\newcommand{\pftwo}[2]{\frac{\partial^2 #1}{\partial #2^2}}
\newcommand{\pfmix}[3]{\frac{\partial^2 #1}{\partial #2\partial #3}}

\newcommand{\abs}[1]{\left| #1\right|}
\newcommand{\tr}[1]{\mathrm{tr}\left[#1\right]}
\newcommand{\vect}[1]{\mathrm{vec}\left(#1\right)}

\DeclareFontFamily{OT1}{pzc}{}
\DeclareFontShape{OT1}{pzc}{m}{it}{<-> s * [1.10] pzcmi7t}{}
\DeclareMathAlphabet{\mathpzc}{OT1}{pzc}{m}{it}

\newcommand{\Order}[1]{\mathcal{O}\lp#1\rp}

\usepackage[pdftex,final,bookmarksnumbered,bookmarksopen=false,breaklinks,colorlinks]{hyperref}
\hypersetup{
	final,
    bookmarksnumbered=true,
    bookmarksopen=true,
    bookmarksopenlevel=0,
    unicode=false,          % non-Latin characters in Acrobat?s bookmarks
    pdftoolbar=true,        % show Acrobat's toolbar?
    pdfmenubar=true,        % show Acrobat's menu?
    pdffitwindow=false,     % window fit to page when opened
    pdftitle={Langevin diffusions on the torus: estimation and applications},    % title
	pdfdisplaydoctitle=true, %display document title instead of file name in title bar
	pdftoolbar=true,
	pdfmenubar=true,
	pdflang={English},
    pdfauthor={Eduardo Garcia-Portugues, Michael Sorensen, Kanti V. Mardia, Thomas Hamelryck},     % author
    pdfsubject={arXiv paper},   % subject of the document
    pdfcreator={Eduardo Garcia-Portugues},   % creator of the document
    pdfproducer={Eduardo Garcia-Portugues}, % producer of the document
    pdfkeywords={Circular data}{Directional statistics}{Likelihood}{Protein structure}{Stochastic differential equation}{Wrapped normal}, % list of keywords
    pdfnewwindow=true,      % links in new window
    breaklinks=true,
    hidelinks,
    linkcolor=black,          % color of internal links (change box color with linkbordercolor)
    citecolor=black,        % color of links to bibliography
    filecolor=magenta,      % color of file links
    urlcolor=cyan           % color of external links
}

\newtheorem{defin}{Definition}

\newtheorem{coro}{Corollary}
\newtheorem{rem}{Remark}
\newtheorem{prop}{Proposition}
\newtheorem{lem}{Lemma}

\allowdisplaybreaks

\newif\ifmain
\maintrue
\ifmain

\else

\fi
\newif\ifsupplement
\supplementtrue

\begin{document}

\ifmain

%-------------- TITLE ----------------%
\title{Langevin diffusions on the torus: estimation and applications}
\setlength{\droptitle}{-1cm}
\predate{}%
\postdate{}%

%-------------- AUTHORS --------------%
\date{}

\author{Eduardo Garc\'ia-Portugu\'es$^{1,2,3,7}$, Michael S\o rensen$^{1}$,\\ Kanti V. Mardia$^{4,5}$, and Thomas Hamelryck$^{2,6}$}

\footnotetext[1]{
Department of Mathematical Sciences, University of Copenhagen (Denmark).}
\footnotetext[2]{
Bioinformatics Centre, Section for Computational and RNA Biology, Department of Biology, University of Copenhagen (Denmark).}
\footnotetext[3]{
Department of Statistics, Carlos III University of Madrid (Spain).}
\footnotetext[4]{
Department of Statistics, University of Leeds (UK).}
\footnotetext[5]{Department of Statistics, University of Oxford (UK).}
\footnotetext[6]{
Image Section, Department of Computer Science, University of Copenhagen (Denmark).}
\footnotetext[7]{Corresponding author. e-mail: \href{mailto:edgarcia@est-econ.uc3m.es}{edgarcia@est-econ.uc3m.es}.}

\maketitle

%-------------------------------------%

\begin{abstract}
We introduce stochastic models for contin\-u\-ous-time evolution of angles and develop their estimation. We focus on studying Langevin diffusions with stationary distributions equal to well-known distributions from directional statistics, since such diffusions can be regarded as toroidal analogues of the Ornstein--Uhlenbeck process. Their likelihood function is a product of transition densities with no analytical expression, but that can be calculated by solving the Fokker--Planck equation numerically through adequate schemes. We propose three approximate likelihoods that are computationally tractable: (\textit{i}) a likelihood based on the stationary distribution; (\textit{ii}) toroidal adaptations of the Euler and Shoji--Ozaki pseudo-likelihoods; (\textit{iii}) a likelihood based on a specific approximation to the transition density of the wrapped normal process. A simulation study compares, in dimensions one and two, the approximate transition densities to the exact ones, and investigates the empirical performance of the approximate likelihoods. Finally, two diffusions are used to model the evolution of the backbone angles of the protein G (PDB identifier 1GB1) during a molecular dynamics simulation. The software package \texttt{sdetorus} implements the estimation methods and applications presented in the paper.
\end{abstract}
\begin{flushleft}
\small\textbf{Keywords:} Circular data; Directional statistics; Likelihood; Protein structure; Stochastic differential equation; Wrapped normal.
\end{flushleft}

%-------------------------------------%
\section{Introduction}
\label{sec:intro}
%-------------------------------------%

Useful proposals of stochastic processes must take into account the particular features of the data that they aim to model. This is so for \textit{toroidal} data, where observations are elements on the torus
$\T^p=[-\pi,\pi)\times\overset{p}{\cdots}\times[-\pi,\pi)$ (with $-\pi$ and $\pi$ identified). Models and
inference for \textit{circular} data ($p=1$) are notably different from the Euclidean
case; see \cite{Mardia2000} or \cite{Jammalamadaka2001} for
a comprehensive description and a review of applications. One of the first continuous-time processes on the circle was proposed by \cite{Kent1975}. It is defined
as the solution to the Stochastic Differential Equation (SDE)
\begin{align}
\rd \Theta_t=\alpha\sin(\mu-\Theta_t)\rd t+\sigma\rd W_t,\label{eq:vMOU}
\end{align}
where $\{W_t\}$ is a Wiener process, $\alpha>0$ is the strength of the drift towards $\mu\in[-\pi,\pi)$, and $\sigma>0$ is the diffusion coefficient. This process, referred to below as the von Mises (vM) process, can be regarded as a circular analogue of the Ornstein--Uhlenbeck (OU) process. The process is attracted to $\mu$ and, in the neighbourhood of $\mu$, the drift is approximately linear. Moreover, the process is ergodic (\textit{i.e.}, it has a unique stationary distribution) and the stationary distribution (abbreviated as sdi) is $\mathrm{vM}\big(\mu,
\frac{2\alpha}{\sigma^2}\big)$. $\mathrm{vM}(\mu,\kappa)$ denotes
the vM distribution with probability density function (pdf)
\begin{align*}
f_{\mathrm{vM}}(\theta;\mu,\kappa):=\frac{e^{\kappa\cos(\theta-\mu)}}{2\pi\Ical_0(\kappa)},\quad \theta,\mu\in[-\pi,\pi),\quad \kappa\geq0,
\end{align*}
with $\Ical_\nu$ being the modified Bessel
function of the first kind and order $\nu$. Despite its similarities with the OU process, the vM process is not as tractable as the former: no analytical expression for its transition probability density (tpd) is known. The vM process has been applied in
mathematical biology \citep{Hill1997,Codling2005}, and related
extensions were studied in physics in the context
of oscillators (see Section 5.3.3 in \cite{Frank2005} and references
therein).\\

The contributions of this paper are two-fold. Firstly, we propose
ergodic diffusions on the torus whose sdis are well-established
distributions from directional statistics. These diffusions can be regarded as
toroidal analogues of the OU process. Specifically, we
introduce several Langevin diffusions, each defined as the wrapping of a $p$-dimensional Euclidean diffusion solving the time-homogeneous SDE
\begin{align}
\rd \bX_t=b(\bX_t)\rd t+\sigma(\bX_t)\rd\bW_t, \label{eq:sde}
\end{align}
where $b:\R^{p}\rightarrow\R^{p}$ is the drift, $\sigma:\R^{p}\rightarrow\R^{p\times p}$ is the diffusion coefficient, and $\bW_t=(W_{t,1},\ldots,W_{t,p})'$ is a vector of $p$ independent standard Wiener processes ($'$ denotes transposition). We provide insights on the wrapping of \eqref{eq:sde} and study the properties of the new diffusions. We give particular emphasis to the Langevin diffusion with Wrapped Normal (WN) sdi, since this is a toroidal OU analogue with more tractable estimation. \\

Secondly, we present estimation procedures for discretely observed
toroidal diffusions. The likelihood function involves the evaluation of the tpd
$p_t(\cdot\, | \,\bx_s)$, the density function of the conditional
distribution of $\bX_{t+s}$ given $\bX_s=\bx_s$. The tpd solves the
Fokker--Planck or Kolmogorov's Forward equation, this is, the Partial Differential Equation (PDE)
\begin{align}
\pf{}{t} p_t(\bx\, | \,\bx_s)=-\sum_{i=1}^p\frac{\partial}{\partial
  x_i} (b_i(\bx)p_t(\bx\,|\,\bx_s)) +\frac{1}{2}\sum_{i,j=1}^p\frac{\partial^2}{\partial x_i\partial x_j}
  (V_{ij}(\bx)p_t(\bx\, | \,\bx_s)), \label{eq:pde}
\end{align}
with $\bx,\bx_s\in\R^p$, $V(\cdot):=\sigma(\cdot)\sigma(\cdot)'$ and initial condition $p_0(\bx\, | \,\bx_s)=\delta(\bx-\bx_s)$ ($\delta(\cdot)$ represents Dirac's delta). This PDE has no explicit solution except for very few particular choices
of $b$ and $V$.  We consider
maximum likelihood estimation based on the numerical solution of \eqref{eq:pde}. This method is computationally costly, but serves as a benchmark to which other computationally more expedient methods can
be compared. A simple solution is to replace the unknown tpd by the
known sdi, hence reducing the problem to
maximum likelihood estimation with independent and identically distributed data, but this is usually inefficient and only allows for the estimation of the parameters appearing in the sdi. We
therefore develop better approximations to the tpd that are relatively
easy to compute. For general diffusions, we introduce toroidal versions of the
Euler and Shoji--Ozaki pseudo-likelihoods. For the WN process, we derive a specific, sdi-correct and
computationally efficient tpd approximation. We investigate the quality of these estimators by calculating
the Kullback--Leibler divergences of the approximating tpds with
respect to the tpd obtained by numerically solving \eqref{eq:pde}. Furthermore, in a simulation study for different discretization steps we compare, in the one- and two-dimensional cases,
the performance of the proposed approximate likelihoods. \\

Next, we describe relevant literature to our contributions. Diffusions featuring trigonometric drifts were presented in
\cite{Kessler1999}, \cite{Larsen2007} and \cite{Sorensen2012},
although these processes are not designed to capture periodicity, but rather to have a bounded interval as their state
space. Wrapped Gaussian processes have been considered by
\cite{Jona-Lasinio2012} in the context of spatial modelling of wave
directions. In a different setting, processes where the time-inhomogeneous drift
$b(t,X_t)$ is a periodic function of time have
been studied by \cite{Dehay2015} and \cite{Dehling2010}. Discrete time
processes on the circle include the circular autoregressive models by
\cite{Breckling1989} and the Markov processes on the circle by
\cite{Wehrly1980}, \cite{Kato2010} and \cite{Yeh2013}. In a
broader perspective, stochastic calculus on manifolds has been
extensively developed, see for example
\cite{Emery1989}, \cite{Stroock2000} and \cite{Hsu2002}. For the case of the torus, a flat and compact manifold, the modelling
challenges do not reside in the curvature of the manifold, but
rather in capturing angular dependencies, a non-trivial and
ubiquitous problem in directional statistics, consequence of the
complex behaviour of rotations on the torus. Finally, we refer to \cite{Rogers2000a},
\cite{Steele2001} and \cite{Oksendal2003} for an exhaustive
introduction to SDEs, and to \cite{Kloeden1992} and \cite{Iacus2008}
for a more applied perspective.\\

The rest of this paper is organized as follows. Section
\ref{sec:ouproc} introduces diffusions on
the torus. Section \ref{sec:est} presents and analyses several estimation
procedures for them, whilst the empirical estimation performance is assessed in a simulation study in Section
\ref{sec:sim}. Section \ref{sec:app} gives an application to modelling
the evolution of protein backbone angles. Conclusions and final comments are given in Section \ref{sec:conc}.

%-------------------------------------%
\section{Toroidal diffusions}
\label{sec:ouproc}
%-------------------------------------%

The state space of a stochastic process $\lrb{\bTheta_t}$ on the torus is $\T^p=[-\pi,\pi)\times\overset{p}{\cdots}\times[-\pi,\pi)$. The space $\R^p$ also plays a relevant role, since $\lrb{\bTheta_t}$ can be regarded as a Euclidean process $\lrb{\bX_t}$ that is wrapped into its principal angles by $\cmod{\cdot}:=((\cdot+\pi)\mod 2\pi)-\pi$. This approach eases the interpretation of crossings through boundaries and motivates the following definition.

\begin{defin}[Toroidal diffusion]
\label{def:tordi}
The stochastic process $\{\bTheta_t\}\subset\T^p$ is said to be a \emph{toroidal diffusion} if it arises as the wrapping $\bTheta_t=\cmod{\bX_t}$ of a diffusion \eqref{eq:sde} such that $b$ and $\sigma$ are $2\pi$-periodic:
\[
b(\bx+2\bk\pi)=b(\bx),\,\sigma(\bx+2\bk\pi)=\sigma(\bx),\,\forall\bk\in\Z^p,\,\forall\bx\in\R^p.
\]
The toroidal diffusion coming from the wrapping of \eqref{eq:sde} is denoted as $\rd \bTheta_t=b(\bTheta_t)\rd t+\sigma(\bTheta_t)\rd \bW_t$.
\end{defin}

The periodicity of $b$ and $\sigma$ are required to make $\{\bTheta_t\}$ a diffusion, since $\{\bTheta_t\}$ can only be Markovian if $\{\bX_t\}$ is non-ergodic in $\R^p$, as the next result shows.

\begin{prop}[Wrapped ergodic diffusion]
\label{prop:condden}
Let $\{\bX_t\}$ be an ergodic diffusion on $\R^p$ with stationary density $\nu$ and tpd $p_t(\cdot\, | \,\bx_s)$. The following statements hold for the wrapped process $\bTheta_t:=\cmod{\bX_t}$:
\begin{enumerate}[label=\textit{\roman*.}, ref=\textit{\roman*}]
\item $\{\bTheta_t\}$ is ergodic on $\T^p$, with stationary density
$\nu^{\mathrm{W}}(\cdot):=\sum_{\bk\in\Z^p} \nu (\cdot+2\pi\bk)$.
\item If $\bX_s$ is distributed with density $\nu$, then the conditional density of $\bTheta_{t+s}\,|\,\bTheta_s=\btheta_s$ is
\begin{align}
p^{\mathrm{W}}_t(\cdot \, | \, \btheta_s) &:= \sum_{\bk,\bm\in\Z^p} p_t(\cdot + 2\bk\pi  \, | \, \btheta_s + 2\bm\pi )w_{\bm}(\btheta_s),\quad w_{\bm}(\cdot):=\frac{\nu(\cdot+2\bm\pi )}{\nu^{\mathrm{W}}(\cdot)}.\label{prop:condden:1}
\end{align}
\item If $\{\bX_t\}$ is time-reversible, \textit{i.e.}, $p_t(\bx\,|\,\by)\nu(\by)=p_t(\by\,|\,\bx)\allowbreak \nu(\bx)$, $\forall\bx,\by\in\R^p$, then $p^{\mathrm{W}}_t(\btheta\,|\,\bphi)\nu^{\mathrm{W}}(\bphi)=p_t^{\mathrm{W}}(\bphi\,|\,\btheta)\allowbreak \nu^{\mathrm{W}}(\btheta)$, $\forall\btheta,\bphi\in\T^p$.
\item The wrapped process is \emph{not} Markovian.
\end{enumerate}
\end{prop}
\begin{proof}
The statements can be easily checked, so we only illustrate the non-Markovianity. Recall that for $t_0<t_1<t_2$ and using the Markovianity of $\lrb{\bX_t}$,
\begin{align*}
\mathbb{P}\{\bTheta_{t_2}\in B \, | &\, \bTheta_{t_1} = \btheta_{t_1}, \, \bTheta_{t_0} =
\btheta_{t_0}\} \\
=&\, \sum_{\bm,\bn\in\Z^p} \Prob{\bTheta_{t_2}\in B \, | \, \bX_{t_1} = \btheta_{t_1} +
	2\bm\pi } \frac{p_{t_1-t_0}(\btheta_{t_1}+2\bm\pi  \, | \, \btheta_{t_0} + 2\bn\pi )}{p^{\mathrm{W}}_{t_1-t_0}(\btheta_{t_1} \, | \, \btheta_{t_0})}w_{\bn}(\btheta_{t_0}).
\end{align*}
This clearly depends on $\btheta_{t_0}$ unless $p_t$ is periodic on both arguments, impossible for a density in $\R^p$.
\end{proof}

Thus, a wrapped ergodic diffusion is \emph{not} a diffusion. In particular, the family of conditional distributions given by \eqref{prop:condden:1} does not define a semi-group of transition operators. The non-Markovianity arises because $\bTheta_{t_2}\,|\,(\bTheta_{t_1}, \,\allowbreak \bTheta_{t_0})$, with $t_2>t_1>t_0$, does not depend only on $\bTheta_{t_1}$ but also on the \emph{winding number} $\mathrm{wind}(\bX_{t_1})\allowbreak:=\lfloor\frac{\bX_{t_1}+\pi}{2\pi}\rfloor\in\Z^p$ of $\bX_{t_1}=\bTheta_{t_1}+2\mathrm{wind}(\bX_{t_1})\pi$, hence the requirement for periodic $b$ and $\sigma$ in Definition \ref{def:tordi}.

\begin{rem}
The density of $\bTheta_{t+s}\,|\,\bTheta_s=\btheta_s$ is remarkably different from the density of $\bTheta_{t+s}\,|\,\bX_s=\btheta_s$, given by $\sum_{\bk\in\Z^p}p_t(\cdot+2\bk\pi\,|\, \bX_s=\btheta_s)$. To make this point clearer, let $\lrb{X_t}$ be the OU process $\rd X_t=\alpha(\mu-X_t)\rd t+\sigma\rd W_t$, $\lrb{\Theta_t}$ its wrapped version with sdi $\sum_{k\in\Z}
\phi_{\sigma/\sqrt{2\alpha}}(\theta\allowbreak -\mu+2k\pi)$ ($\phi_\sigma$ is the pdf of a $\mathcal{N}(0,\sigma^2)$), and assume $X_s\sim\mathcal{N}\big(\mu,\frac{\sigma^2}{2\alpha}\big)$. Then:
\begin{enumerate}[label=\roman*., ref=\roman*]
\item The density of $X_{t+s}\,|\,X_s=\theta_s$ is $\phi_{\sigma_t}(\cdot-\mu_t)$, with $\mu_t:=\mu+(\theta_s-\mu)e^{-\alpha t}$ and $\sigma_t^2:=\frac{\sigma^2}{2\alpha}(1-e^{-2\alpha t})$. This is the usual tpd of the OU process.
\item The density of $X_{t+s}\,|\,\Theta_s=\theta_s$ is $\sum_{m\in\Z}\phi_{\sigma_t}(\cdot-\mu_t^m)\allowbreak w_m(\theta_s)$, where $\mu^m_t := \mu + (\theta_s + 2\pi m - \mu)e^{-\alpha t}$ and $w_m(\theta_s) = \frac{\phi_{\sigma/\sqrt{2\alpha}}(\theta_s-\mu+2\pi m)}{\sum_{k\in\Z} \phi_{\sigma/\sqrt{2\alpha}}(\theta_s-\mu+2k\pi)}$.
\item The density of $\Theta_{t+s}\,|\,\Theta_s=\theta_s$ is $p_t^{\mathrm{W}}(\cdot\,|\, \Theta_s=\theta_s)=\sum_{k,m\in\Z} \phi_{\sigma_t} \lrp{\cdot - \mu^m_t  + 2k\pi }w_m(\theta_s)$. This circular density can exhibit two modes describing the drift of $\{\Theta_t\}$ towards $\mu$ whenever $\theta_s$ and $\mu$ are antipodal.
\item The density of $\Theta_{t+s}\,|\,X_s=\theta_s$ is $\sum_{k\in\Z} \phi_{\sigma_t} (\cdot - \mu_t  +\allowbreak 2k\pi)$, which is unimodal. Whenever the circular shortest distance between $\theta_s$ and $\mu$ happens across the boundary, this circular density pushes the probability mass in the opposite direction. \label{rem:1}
\end{enumerate}
\end{rem}

\begin{rem}
\cite{Liu2013} stated a similar density to \ref{rem:1} above, with $2k\pi e^{-t}$ instead of $2\pi k$, as the ``tpd function of the OU process on the circle'' and proved it satisfied the Chapman--Kolmogorov equation\footnote{Note that $-(x_2-x_1e^{-(t_2-t_1)}+2k\pi e^{-t_2})^2$ should be in the exponential's denominator of \cite{Liu2013}'s (15) and \nolinebreak[4](16).}. That density is not circular (it has a time-shrinking period $2k\pi e^{-t}$).
\end{rem}

The rest of the section is devoted to the introduction and analysis of notable toroidal diffusions.

%-------------------------------------%
\subsection{Langevin toroidal diffusions}
%-------------------------------------%

Let $f$ be a pdf over $\R^p$. The so-called \textit{Langevin diffusions} are a family of multivariate diffusions of the form \eqref{eq:sde}, where the entries of $b$ are given by
\begin{align}
b_i(\bx)=\frac{1}{2}\sum_{j=1}^pV_{ij}(\bx)\pf{}{x_j}\log f(\bx) +\det V(\bx)^\frac{1}{2}\sum_{j=1}^p\pf{}{x_j}\lrp{V_{ij}(\bx)\det V(\bx)^{-\frac{1}{2}}},\label{eq:lange}
\end{align}
with $i=1,\ldots,p$. The most important property of these diffusions is that, under mild regularity conditions on $f$ and $\sigma$, they are ergodic with stationary density $f$. This is particularly convenient since one of the first steps in modelling a given trajectory is to compare its empirical distribution with the sdi of the candidate diffusion model. Remarkably, the family of Langevin diffusions characterizes the family of ergodic diffusions with a given sdi that are time-reversible. The result is due to \cite{Kolmogoroff1937} and was later extended by \cite{Kent1978} using symmetric diffusions on manifolds (see Theorems 4.2 and 6.1 \textit{ibid}). In particular, the OU process is identified as the unique time-reversible diffusion with Gaussian sdi and constant diffusion coefficient. This characterization is key for constructing analogues of the OU process in $\T^p$ by means of Langevin diffusions driven by Gaussian-like toroidal distributions. \\

The construction of Langevin toroidal diffusions is achieved by wrappings of Langevin diffusions, where now $f$ is a toroidal density, that is, $\int_{\T^p}f(\btheta)\rd\btheta=1$ and $f(\btheta+2\bk\pi)=f(\btheta)$, $\forall\btheta\in\T^p,\,\bk\in\Z^p$.

\begin{prop}
\label{prop:lang}
Assume $\{\bTheta_t\}$ is obtained from the wrapping of a Langevin diffusion $\{\bX_t\}$ with drift \eqref{eq:lange}, given by a strictly positive toroidal density $f$. Assume that the second derivatives of both $f$ and the entries of $V$ are H\"{o}lder continuous, and that $V$ is $2\pi$-periodic. Then, for the given $V$, $\{\bTheta_t\}$ is the unique toroidal time-reversi\-ble diffusion that is ergodic with stationary density $f$ and squared diffusion coefficient $V$.
\end{prop}
\begin{proof}
We provide a sketch. The time-reversibility with equilibrium density $f$ follows from
Theorem 10.1 in \cite{Kent1978} using the compactness (makes $\{\bTheta_t\}$ conservative), flatness, and global coordinates of $\mathbb{T}^p$. The equilibrium distribution is also the
(unique) sdi, so $\{\bTheta_t\}$ is ergodic. To show the uniqueness, note that by Theorem 6.1 \textit{ibid} a time-reversible diffusion must have an equilibrium density $u$ and be
$u$-symmetric, where necessarily $u=f$. By Theorem 4.2 \textit{ibid} (and its
proof) the only way a diffusion with a given $V$ can be $f$-symmetric is if its drift is \eqref{eq:lange}.
\end{proof}
As a consequence, any time-reversible toroidal diffusion with stationary density $f$ and $V(\bx)=\bSigma$ is of the form
\begin{align*}
\rd \bTheta_t=\frac{1}{2}\bSigma\nabla\log f(\bTheta_t)\rd t+\bSigma^\frac{1}{2}\rd\bW_t.
\end{align*}
The rest of the paper focuses on diffusions of this form.

%-------------------------------------%
\subsection{Analogues of the Ornstein--Uhlenbeck process}
%-------------------------------------%

The vM process can be considered as the circular analogue of the OU process \citep{Kent1975}. Two arguments support this claim: (\textit{i}) the vM process is the unique time-reversible diffusion with vM sdi and constant diffusion coefficient; (\textit{ii}) the vM distribution is usually regarded as the Gaussian circular analogue due to important Gaussian-like characterizations \citep[Section 2.2.4]{Jammalamadaka2001}. However, it is worth to note that a similar argument to (\textit{ii}) holds for the WN: this distribution exhibits certain similarities with the Gaussian (\textit{ibid}, Section 2.2.6) and, contrary to the vM distribution, it appears in Gaussian-related limit laws such as the wrapped version of the central limit theorem \citep[Section 4.3.2]{Mardia1972}. \\

In this section we investigate the main properties of the Langevin diffusions driven by the multivariate versions of the vM and WN distributions. In addition, we consider two appealing extensions driven by more flexible sdis: the symmetric circular distribution of \cite{Jones2005} and mixtures of (independent) vM distributions. These processes are later employed in Section \ref{sec:sim}.

%-------------------------------------%
\subsubsection{Multivariate von Mises}
%-------------------------------------%

The multivariate extension of the vM distribution is not immediate: several competing alternatives are described in the literature, see \cite{Mardia2012a} for a review focused on the bivariate case. Among the available proposals, we chose the Multivariate von Mises (MvM) with \textit{sine interaction} \citep{Mardia2008} due to its pleasant modelling properties: simple unimodal characterization, ability of capturing positive/negative dependence within the same density formulation, and vM conditional distributions. The $\mathrm{MvM}(\bmu,\bkappa,\bLambda)$ pdf is
\begin{align*}
f_{\mathrm{MvM}}(\btheta;\bmu,\bkappa,\bLambda):=T(\bkappa,\bLambda)^{-1}\exp\bigg\{\bkappa' \cos(\btheta-\bmu)+\frac{1}{2}\sin(\btheta-\bmu)'\bLambda \sin(\btheta-\bmu)\bigg\},
\end{align*}
where the trigonometric functions are understood as entry-wise operators, $\bkappa\geq0$, $\bLambda$ is a symmetric matrix with zero diagonal, and $T(\bkappa,\bLambda)$ is the normalizing constant. If $\bLambda=\mathbf 0$, then the MvM distribution is the product of independent vM, and hence $T(\bkappa,\mathbf 0)=(2\pi)^p\prod_{j=1}^p\Ical_0(\kappa_j)$. A sufficient condition for unimodality is that $\mathbf{P}:=\diag{\bkappa}-\bLambda$ is positive definite \citep{Mardia2014}, a result related to the fact that, for large concentrations $\bkappa$, $\mathrm{MvM}(\bmu,\bkappa,\bLambda)\approx \mathcal{N}_p(\bmu,\mathbf{P}^{-1})$. The operator $\mathrm{diag}(\cdot)$ denotes either the diagonal extraction or the diagonal matrix construction, depending on its argument.\\

The non-linear dependence structure of the MvM distribution forces $\bSigma$ in the associated Langevin diffusion to be isotropic (\textit{i.e.}, $\bSigma=\sigma^2\bI$) if separability between the drift and diffusion coefficients is desired. We opted to preserve separability and to generalize \eqref{eq:vMOU} by having a $\mathrm{MvM}\big(\bmu,\frac{2\ba}{\sigma^2},\frac{2\bA^*}{\sigma^2}\big)$ sdi:
\begin{align*}
\rd \bTheta_t=\big[\ba\circ \sin(\bmu-\bTheta_t) -(\bA^* \sin(\bmu-\bTheta_t))\circ \cos(\bmu-\bTheta_t)\big]\rd t+\sigma\rd \bW_t,
\end{align*}
where $\circ$ denotes the element-wise product of matrices, $\ba:=\diag{\bA}$, $\bA^*:=\diag{\ba}-\bA$, and $\bA$ is a positive definite matrix. The equilibrium points of drift are located at $\bmu+\bk_0\pi$, with $\bk_0\in\lrb{-1,0,1}^p$ (we assume implicit wrapping by $\mathrm{cmod}$ in the sums of angles in this section), and are unstable if any component is antipodal, this is, unless $\bk_0=\zero$ (see Figures \ref{fig:ld} and \ref{fig:vecdf}). The drift is approximately linear in a neighbourhood of $\bmu$ and has Jacobian $-\bA$. For the unstable points, the drift has Jacobian $-\bA\circ(\bs\bs')$, with $\bs=\cos(\bk_0\pi)$ a vector of signs. In the circular case, the maximal drifts (in absolute value) towards $\mu$ are placed at $\mu\pm\frac{\pi}{2}$ (see Figure \ref{fig:ld}). For the general case, the maximal \textit{marginal} drifts for the $j$-th component happen at $\mu_j-\tan^{-1}\big(A_{jj}\big[\sum_{k\neq j}A_{jk}\sin(\mu_k-\theta_k)\big]^{-1}\big)+k_0\pi$, $k_0\in\lrb{-1,0,1}$.

\begin{figure}[H]
\centering
\includegraphics[width=\textwidth]{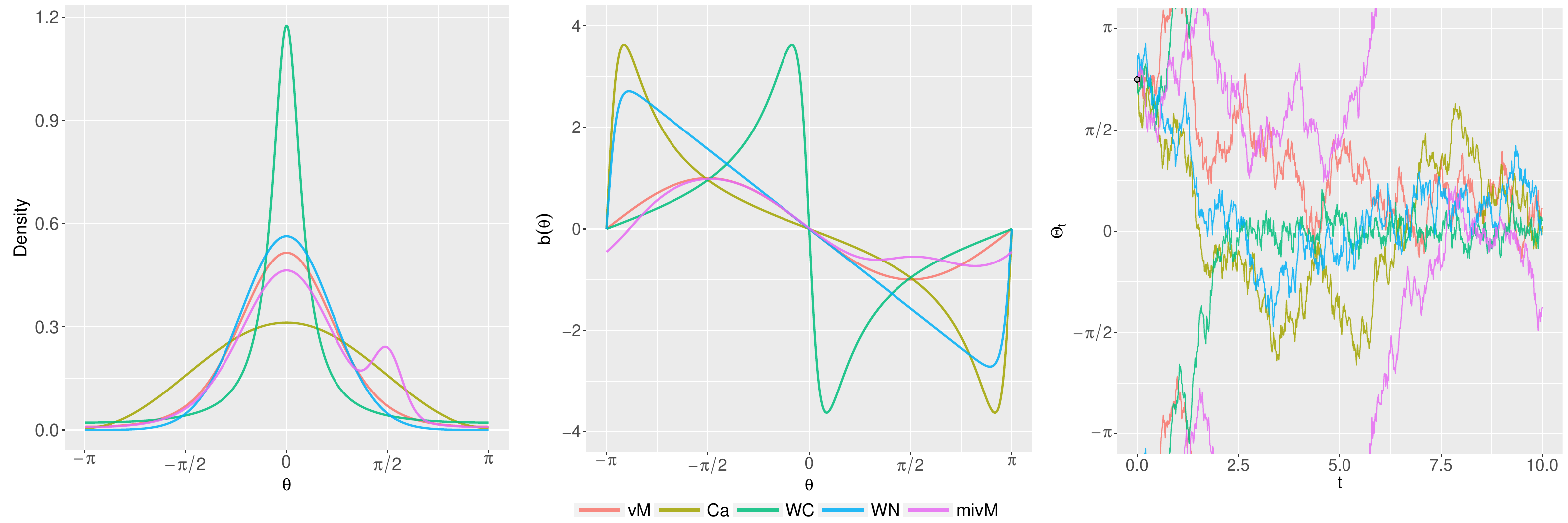}
\caption{\small Stationary pdfs (left panel), drifts (center), and sample trajectories (right) of the one-dimensional vM, WN, WC, Ca, and mivM diffusions. The parameters for the unimodal diffusions are $\mu=0$ and $\alpha=\sigma=1$. The parameters for mivM are $\ba=(1,10)$, $\bmu=(0,\frac{\pi}{2})$, $\bp=(0.9, 0.1)$, and $\sigma=1$. The sample trajectories are simulated in the interval $[0,10]$ with initial point $\frac{3\pi}{4}$. \label{fig:ld}}
\end{figure}

%-------------------------------------%
\subsubsection{Wrapped normal}
%-------------------------------------%

The pdf of a (multivariate) wrapped normal, $\mathrm{WN}\lrp{\bmu,\bSigma}$, is given by $f_{\mathrm{WN}}(\btheta;\bmu,\bSigma):=\sum_{\bk\in\Z^p}\phi_{\bSigma}(\btheta-\bmu+2\bk\pi)$, with $\bmu\in\T^p$, $\bSigma$ a covariance matrix, and $\phi_{\bSigma}$ the pdf of a $\mathcal{N}(\zero,\bSigma)$. For the sake of clarity of exposition, we first introduce the circular case and then the multivariate extension. Using the OU parametrization, the circular WN process with $\mathrm{WN}\big(\mu,\frac{\sigma^2}{2\alpha}\big)$ sdi is defined as
\begin{align}
\rd \Theta_t=\bigg[\alpha\sum_{k\in\Z}(\mu-\Theta_t-2k\pi)w_k(\Theta_t)\bigg]\rd t+\sigma\rd W_t,\quad w_k(\theta)=\frac{ \phi_{\sigma/\sqrt{2\alpha}}\lrp{\theta-\mu+2k\pi}}{\sum_{m\in\Z} \phi_{\sigma/\sqrt{2\alpha}}\lrp{\theta-\mu+2m\pi}}.\label{eq:wn1}
\end{align}
Despite the similar shape of the vM and WN densities in the main bulk of the probability, their behaviour is substantially different at antipodality, a fact strengthened in log scale. The WN process drift is a smoothed ``sawtooth wave'' that has negative slope at $\mu$ and crosses the $x$-axis at $\mu+k\pi$, $k\in\{-1,0,1\}$. Hence, the drift behaves almost linearly in a neighbourhood of $\mu$ (equilibrium point, stable) and rapidly decays to pass across $\mu\pm\pi$ (equilibrium point, unstable). This neighbourhood is larger than for the vM process. There is no separability between $\alpha$ and $\sigma$ and both alter the drift non-trivially. For example, the drift maxima are implicitly given by
$\sum_{k\in\Z}k^2w_k(\theta)-\big[\sum_{k\in\Z}kw_k(\theta)\big]^2=\frac{\sigma^2}{8\alpha\pi^2}$,
and vary from $\mu\pm\pi$ (if $\frac{\sigma^2}{2\alpha}\to0$, the sdi is degenerate at $\mu$) to $\mu\pm\frac{\pi}{2}$ (if $\frac{\sigma^2}{2\alpha}\to\infty$, the sdi is uniform and the drift is null). Thereby, the maximum drifts always happen closer to antipodality than in the vM process (see Figure \ref{fig:ld}).
The slopes of the drift at $\mu$ and $\mu\pm\pi$ are $-\alpha+a_1(\alpha,\sigma)$ and $a_2(\alpha,\sigma)$, respectively, where
\begin{align*}
0\leq a_1(\alpha,\sigma):=\frac{8\pi^2\alpha^2}{\sigma^2}\sum_{k\in\Z}k^2w_k(\mu)\leq\alpha,\quad 0\leq a_2(\alpha,\sigma):=-\alpha+\frac{2\pi^2\alpha^2}{\sigma^2}\bigg[4\sum_{k\in\Z}k^2w_{k}(\mu+\pi)-1\bigg].
\end{align*}
The lower and upper bounds for $a_1(\alpha,\sigma)$ (respectively, $a_2(\alpha,\sigma)$) are attained, with $\alpha$ fixed, when $\sigma\to0$ ($\sigma\to\infty$) and $\sigma\to\infty$ ($\sigma\to0$), respectively. \\

The multivariate extension of \eqref{eq:wn1} is the diffusion
\begin{align}
\rd \bTheta_t=&\,\bigg[\bA\sum_{\bk\in\Z^p}\lrp{\bmu-\bTheta_t-2\bk\pi} w_{\bk}(\bTheta_t)\bigg]\rd t +\bSigma^\frac{1}{2}\rd \bW_t,\label{eq:wnp}\\
w_{\bk}(\btheta)=&\,\frac{ \phi_{\frac{1}{2}\bA^{-1}\bSigma}(\btheta-\bmu+2\bk\pi)}{\sum_{\bm\in\Z^p} \phi_{\frac{1}{2}\bA^{-1}\bSigma}(\btheta-\bmu+2\bm\pi)}.\nonumber
\end{align}
This diffusion has $\mathrm{WN}\lrp{\bmu,\frac{1}{2}\bA^{-1}\bSigma}$ sdi, provided that $\bA$ is invertible and such that $\bA^{-1}\bSigma$ is a covariance matrix. The drift is null at $\bmu+\bk_0\pi$, with $\bk_0\in\lrb{-1,0,1}^p$, since $\sum_{\bk\in\Z^p}(2\bk+\bk_0) w_{\bk}(\bmu+\bk_0\pi)=\zero$ due to the symmetry of $w_{\bk}(\bmu)$ as a function of $\bk\in\Z^p$. Properties similar to the circular case can be obtained using that $\nabla w_{\bk}(\btheta)=4\pi\bSigma^{-1}\bA  w_{\bk}(\btheta)\lrc{\sum_{\bm\in\Z^p}\bm w_{\bm}(\btheta)-\bk}$. For instance, the Jacobian of the drift at $\bmu$ is $-\bA+8\pi^2\bA\big[\sum_{\bk\in\Z^p}\bk\bk'w_{\bk}(\bmu)\big]\allowbreak\bA'\bSigma^{-1}$.\\

The vector field of the drift has a characteristic tessellated structure that, in the two-dimensional case, is formed by hexagonal-like tiles (see Figure \ref{fig:vecdf}). $\bSigma$ alters the tessellation that binds the drifts $\bA(\bmu-\btheta-2\bk\pi)$ by modifying $\{w_{\bk}(\btheta):\bk\in\Z^p\}$. This set is the distribution of the winding numbers of $\bX\sim\mathcal{N}(\bmu,\frac{1}{2}\bA^{-1}\bSigma)$, since
$\mathbb{P}\{\mathrm{wind}(\bX)=\bk\,|\,\cmod{\bX}=\btheta\}=w_\bk(\btheta)$ and satisfies that $\arg\max_{\bk\in\Z^p}w_{\bk}(\btheta)=\mathrm{wind}(\bmu-\btheta)$. Under isotropy (\textit{i.e.} $\bSigma=\sigma^2\bI$), the larger (respectively, smaller) $\sigma$, the more spread (concentrated) the distribution of winding numbers is, resulting in flat (peaked) drifts with smooth (rough) transitions in the limits defining the tessellation.

%-------------------------------------%
\subsubsection{\texorpdfstring{\cite{Jones2005}'s circular distribution}{Jones and Pewsey (2005)'s circular distribution}}
%-------------------------------------%

The \cite{Jones2005} distribution, $\mathrm{JP}(\mu,\kappa,\psi)$, is a tractable family of symmetric and unimodal circular distributions that contains the Wrapped Cauchy (WC, $\psi=-1$), Cardioid (Ca, $\psi=1$), and von Mises ($\psi\to0$) distributions. Its pdf is
$f_{\mathrm{JP}}(\theta;\mu,\kappa,\psi):=(2\pi \allowbreak P_{1/\psi}(\cosh(\kappa\psi)))^{-1}(\cosh(\kappa\psi)\allowbreak+\allowbreak\sinh(\kappa\psi)\allowbreak\cos(\theta-\mu))^{1/\psi}$, with $\mu\in[-\pi,\pi)$, $\kappa\geq0$, $\psi\in\R$, and $P_\nu$ the Legendre function of the first kind and order $\nu$. \\

The diffusion with $\mathrm{JP}\big(\mu,\frac{2\alpha}{\sigma^2},\psi\sigma^2\big)$ sdi, parametrised to yield \eqref{eq:vMOU} as a particular case, is
\[
\rd \Theta_t=\frac{\sinh(2\alpha\psi)\sin(\mu-\Theta_t)}{\psi(\cosh(2\alpha\psi)+\sinh(2\alpha\psi)\cos(\mu-\Theta_t))}
\rd t+\sigma\rd W_t.
\]
The maximal drifts, located at $\mu\pm(\frac{\pi}{2}+\allowbreak\sin^{-1}(\tanh(2\alpha\psi)))$, are closer to the equilibrium mean $\mu$ when $\psi<0$ and to the antipodal mean when $\psi>0$. The slope of the drift at $\mu$ is
$\frac{e^{4\alpha\psi}-1}{4\psi}$. At $\mu\pm\pi$, it is $\frac{e^{-4\alpha\psi}-1}{4\psi}$. This relates to the fact that the drifts with $\psi<0$ equal the ones with $\psi>0$, once translated by $\pm\pi$ and reflected around $\mu$. Hence, whilst the WC diffusion features a drift attracting the process towards a tight neighbourhood around $\mu$, the Ca diffusion repulses the process from $\mu\pm\pi$ and weakly attracts it towards $\mu$ (see trajectories and drifts in Figure \ref{fig:ld}).

\begin{figure}[H]
\centering
\includegraphics[width=\textwidth]{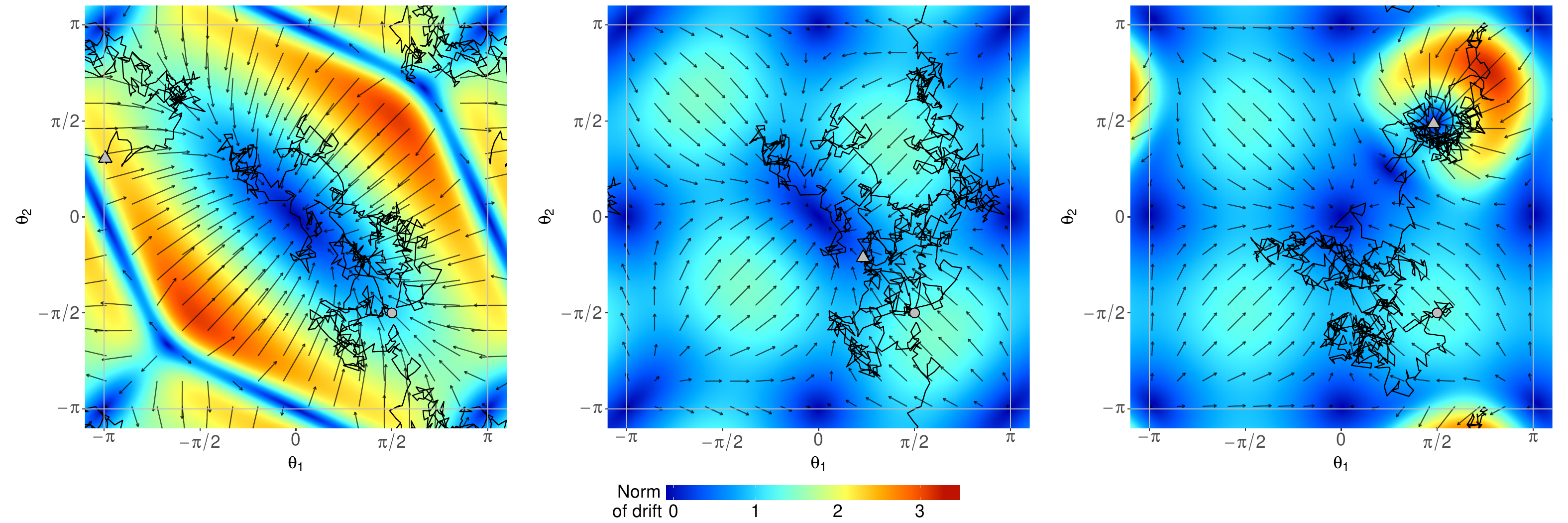}
\caption{\small Vector fields of the drift for the two-dimensional diffusions WN (left panel), MvM (center), and mivM (right). The vector field is coloured by the Euclidean norm of the drift. Parameter $\sigma=1.5$ is common to all panels and $\bmu=(0,0)$ is used for first and second. For MvM and WN, $\bA=\lrp{1,0.5;0.5,1}$ and $\bSigma=\sigma^2\bI$ are considered. For mivM, $\bM=\lrp{0,0;\frac{\pi}{2},\frac{\pi}{2}}$, $\bA=\lrp{1,1;4,4}$, and $\bp=(0.8,0.2)$. A characteristic trajectory starting at $(-\frac{\pi}{2},-\frac{\pi}{2})$ and running in the time interval $[0,5]$ is drawn, with round and triangular facets denoting the start and end of the simulated path, respectively. \label{fig:vecdf}}
\end{figure}

%-------------------------------------%
\subsubsection{Mixtures of independent von Mises}
%-------------------------------------%

The density of an $m$-mixture of independent von Mises distributions, $\mathrm{mivM}(\bM,\bK,\bp)$, is given by $f_\mathrm{mivM}(\btheta;\bM,\allowbreak\bK,\bp)\allowbreak:=\sum_{j=1}^mp_jf_\mathrm{MvM}(\btheta;\bmu_j,\bkappa_j,\mathbf 0)$, with $\bM:=(\bmu_1,\allowbreak\ldots,\allowbreak\bmu_m)'$, $\bK:=(\bkappa_1,\ldots,\bkappa_m)'$, $\bp:=(p_1,\ldots,p_m)'$, and $p_j\geq0$, $j=1,\ldots,m$ and $\sum_{j=1}^mp_j=1$. The mivM distribution is a highly flexible tool for modelling multimodal and skewed circular data \citep{Banerjee2005}, and has tractability as a key advantage: the normalizing constant is known and estimation by the Expectation-Maximization (EM) algorithm is relatively easy. Setting $\bA=(\ba_1,\ldots,\ba_m)'$ and $\bkappa_j=\nolinebreak[4]\frac{2\ba_j}{\sigma^2}$,
\begin{align}
\rd \bTheta_t=\bigg[\sum_{j=1}^m\ba_j\circ \sin(\bmu_j-\bTheta_t)v_j(\bTheta_t)\bigg]\rd t +\sigma\rd \bW_t,\quad v_j(\btheta)=\frac{p_jf_{\mathrm{MvM}}\big(\btheta;\bmu_j,\frac{2}{\sigma^2}\ba_j,\mathbf 0\big)}{\sum_{l=1}^mp_lf_{\mathrm{MvM}}\big(\btheta;\bmu_l,\frac{2}{\sigma^2}\ba_l,\mathbf 0\big)},\label{eq:mixvm}
\end{align}
has $\mathrm{mivM}\big(\bM,\frac{2\bA}{\sigma^2},\bp)$ sdi. The mivM process drift is a weighted average of the corresponding component drifts, whose weights are the posterior probabilities of drawing $\bTheta_t$ from the mixture components of the sdi. The drift behaves locally around $\bmu_j$ as $\ba_j\circ\sin(\btheta-\bmu_j)v_j(\bmu_j)+\bb_j$, with $\bb_j=\sum_{k\neq j}\ba_k\circ \sin(\bmu_k-\bmu_j)v_k(\bmu_j)$ (Figures \ref{fig:ld} and \ref{fig:vecdf}). Then, $\bmu_j$ is only an \textit{asymptotic} equilibrium point for $\sigma\to0$, since $\lim_{\sigma\to0}v_k(\bmu_j)=\delta_{jk}$. The larger $\sigma$, the smoother the binding of the component drifts is.

%-------------------------------------%
\section{Estimation for toroidal diffusions}
\label{sec:est}
%-------------------------------------%

We focus now on the estimation of the vector parameter $\blambda$ of a toroidal diffusion
\begin{align}
\rd \bTheta_t=b(\bTheta_t;\blambda)\rd t+\sigma(\bTheta_t;\blambda)\rd \bW_t,\label{eq:param}
\end{align}
when the data are observations at discrete time points,
$\{\bTheta_{\Delta i}\}_{i=0}^N$. For simplicity, we assume that the
time points are equidistant in the time interval $[0,T]$,
$T=N\Delta$. The Maximum Likelihood\nopagebreak[4] Estimator (MLE) for $\blambda\in\Lambda$ is
given by
\[
\hat\blambda_{\mathrm{MLE}}:=\arg\max_{\blambda\in\Lambda}
l(\blambda;\{\bTheta_{\Delta i}\}_{i=0}^N),
\]
where, using the Markovianity of \eqref{eq:param}, the log-likelihood is given by
\begin{align}
l\lrp{\blambda;\{\bTheta_{\Delta i}\}_{i=0}^N}=\log
  p(\bTheta_{0};\blambda) +\sum_{i=1}^N\log p_{\Delta}(\bTheta_{\Delta
  i}\,|\,\bTheta_{\Delta (i-1)};\blambda).\label{eq:ll}
\end{align}
Here $p_{\Delta}(\cdot\,|\,\cdot;\blambda)$ is the tpd of \eqref{eq:param}. The first term in \eqref{eq:ll} is often disregarded or set to the sdi of \eqref{eq:param}. Maximum likelihood estimation is, under weak regularity conditions, consistent
and asymptotically efficient when $N\to\infty$ with fixed $\Delta$
\citep{Dacunha-Castelle1986}, or when $\Delta\to0$ and $T\to\infty$
\citep{Sorensen2008}. However, it can rarely be readily
performed, as usually no explicit expression for the tpd
exists and this tpd is only given implicitly as the solution to \eqref{eq:pde} on $\T^p$.\\

In the following we present and analyse several estimation strategies
to circumvent the unavailability of the tpd when dealing with toroidal
diffusions. All these methods rely on an \textit{approximate likelihood
function}, where the unknown tpd is replaced by an approximation. For
the sake of brevity, we suppress the, implicitly
assumed, dependence on $\blambda$ in the notation.

%-------------------------------------%
\subsection{Estimation based on the stationary distribution}
%-------------------------------------%

The simplest approximate likelihood function is obtained by replacing
the tpd by the stationary density of \eqref{eq:param}. Usually,
the sdi depends only on a function $\blambda^\nu$ of $\blambda$.
For instance, for the WN process, $\blambda=(\bA,\bmu,\bSigma)$ and
$\blambda^\nu=(\bmu,\frac{1}{2}\bA^{-1}\bSigma)$. Therefore, we denote the
stationary density by $\nu(\cdot;\blambda^\nu)$ and state the \textit{Stationary} MLE (SMLE) of $\blambda^\nu$\nolinebreak[4] as
\begin{align}
\hat\blambda^\nu_{\mathrm{SMLE}}:=\arg\max_{\blambda^\nu\in\Lambda^\nu}
\sum_{i=0}^N\log\nu(\bTheta_i;\blambda^\nu). \label{eq:smle}
\end{align}
For the vM process, SMLE is semi-explicit \cite[page
198]{Mardia2000}. The JP distribution has implicit SMLE and is discussed in \citet[Section 3]{Jones2005}. Effective estimation in MvM distributions involves
pseudo-likelihood \citep{Mardia2008}. Finally, inference for mivM distributions
can be carried out by the EM algorithm \citep{Banerjee2005}. The simple
SMLE is of interest for three reasons: (\textit{i}) for stationary ergodic
processes, it is consistent for $\blambda^\nu$ as $N\to\infty$ for
fixed $\Delta$
\citep{Kessler2000}; (\textit{ii}) $\hat\blambda^\nu_{\mathrm{SMLE}}$ can
be used as a sensible starting value in the optimization routines
of more sophisticated procedures; (\textit{iii}) it can be
supplemented by estimators of the rest of $\blambda$ (see \cite{Bibby2001}, for example).\\

When the unidentifiability of $\blambda$ by SMLE involves the
diffusion matrix $\bSigma$, an estimator of $\blambda$ can be obtained by an estimator of $\bSigma$
that is unrelated to the SMLE. Conditionally on $\bTheta_{\Delta(i-1)}$, $\bTheta_{\Delta i}$
is approximately distributed as $\mathrm{WN}(\bTheta_{\Delta
  (i-1)},\Delta\bSigma)$ when $\Delta\approx0$ (high frequency
observations). This, plus the high concentration of such WN distribution (see Remark
\ref{rem:wn} below), gives
\begin{align*}
p_{\Delta}(\bTheta_{\Delta i}\,|\,\bTheta_{\Delta (i-1)})\approx
  f_\mathrm{WN}(\bTheta_{\Delta i};\bTheta_{\Delta
  (i-1)},\Delta\bSigma) \approx
  \phi_{\Delta\bSigma}\lrp{\cmod{\bTheta_{\Delta i}-\bTheta_{\Delta
  (i-1)}}}.
\end{align*}
Thus, an approximate MLE of $\bSigma$ is
\begin{align}
\hat\bSigma_\mathrm{HF}:=\frac{1}{N\Delta}\sum_{i=1}^N\cmod{\bTheta_{\Delta i}-\bTheta_{\Delta (i-1)}} \cmod{\bTheta_{\Delta i}-\bTheta_{\Delta (i-1)}}'.\label{eq:sigma}
\end{align}
Under isotropy,
$\hat\sigma_\mathrm{HF}^2:=p^{-1}\mathrm{tr}[\hat\bSigma_\mathrm{HF}]$. The
Euclidean counterpart of \eqref{eq:sigma} is well-known to be a
consistent estimator of $\bSigma$ as $\Delta \rightarrow 0$ (for fixed
$T$) due to the convergence in probability to the quadratic\nopagebreak[4] variation. The
consistency for $\hat\bSigma_\mathrm{HF}$ follows easily from this
result.\\

The estimator \eqref{eq:sigma} gives a practical method to disentangle
the unidentifiability inherent to SMLE.  We illustrate this with the WN process. The SMLE $(\hat{\bmu},\hat \bS)_\mathrm{SMLE}$ for
$(\bmu,\bS)$, where $\bS = \frac{1}{2}\bA^{-1}\bSigma$, can be found by optimizing \eqref{eq:smle}. The circular means, $\hat \bmu_c:=\mathrm{atan2}\big(\sum_{i=1}^N\sin(\bTheta_{i\Delta}),\allowbreak \sum_{i=1}^N\cos(\bTheta_{i\Delta})\big)$, and the
high-concentration estimate of $\bS$,
$\frac{1}{N}\sum_{i=1}^N\cmod{\bTheta_{i\Delta}-\hat
  \bmu_c} \allowbreak\cmod{\bTheta_{i\Delta}-\hat \bmu_c}'$, can be used as starting values. $(\hat{\bmu},\hat \bS)_\mathrm{SMLE}$ and
\eqref{eq:sigma} give $\hat\bA:=\frac{1}{2}\hat\bSigma_\mathrm{HF}\bS_\mathrm{SMLE}^{-1}$,
resulting in  $\hat{\blambda}=(\hat{\bA}, \hat{\bmu}_\mathrm{SMLE},
\hat{\bSigma}_\mathrm{HF})$. Similar approaches can be followed for
the rest of the diffusions presented in Section \ref{sec:ouproc}.

%-------------------------------------%
\subsection{Adapted Euler and Shoji--Ozaki pseudo-likelihoods}
%-------------------------------------%

The well-known Euler pseudo-likelihood can be adapted for toroidal
diffusions with minor changes. The Euler scheme arises as the
first order discretization of the process, where the drift
and diffusion coefficient are approximated constantly. After wrapping,
the scheme becomes
\begin{align*}
\bTheta_{\Delta i}=\mathrm{cmod}\big(\bTheta_{\Delta(i-1)}+b(\bTheta_{\Delta(i-1)})\Delta +\sqrt{\Delta}\sigma(\bTheta_{\Delta(i-1)})\mathbf{Z}^i\big),
\end{align*}
where $\mathbf{Z}^i\sim\mathcal{N}(\mathbf{0},\bI)$, $i=1,\ldots,N$. The wrapping
yields the Euler pseudo-tpd
\[
p^\mathrm{E}_{\Delta}(\btheta\,|\,\bphi):=
f_\mathrm{WN}\lrp{\btheta;\bphi+b(\bphi)\Delta,V(\bphi)\Delta},\quad \btheta,\,\bphi\in\T^p.
\]
When $\Delta\to\infty$, the Euler pseudo-tpd converges to the uniform
distribution in $\T^p$ by spreading its probability mass whilst the mean moves along the wrapped line $\{\cmod{\bphi
+ b(\bphi)\Delta}:\Delta>0\}$. The Euler
pseudo-likelihood is obtained from \eqref{eq:ll} by replacing the tpd
by the Euler pseudo-tpd. \\

The Shoji--Ozaki \citep{Shoji1998} scheme uses a linear approximation for the drift and assumes
the diffusion coefficient constant between observation times: for $t\in[s, s + \Delta)$, $b(\bX_t)\approx b(\bX_s)+\bJ_s(\bX_t-\bX_s)$, where
$\bJ_s=J(\bX_s)$ denotes the Jacobian of $b$ at
$\bX_s$. This gives the linear
approximating SDE
\[
\rd \bX_t=(b(\bX_s)+\bJ_s(\bX_t-\bX_s))\rd t+\bsigma_s\rd \bW_t,\quad
t\in[s,s+\Delta).
\]
Conditionally on $\bX_s$, this is a multivariate OU
process. Hence, $\bX_t\,|\,\bX_s\sim\mathcal{N}(\bmu_t,\bGamma_t)$, with $\bmu_t:=\bJ_s^{-1}(\exp\{\bJ_s(t-s)\}-\bI)b(\bX_s)$,
$\bGamma_t:=\int_s^t\exp\{\bJ_s(t-u)\}\bV_s\exp\{\bJ_s'(t-u)\}\rd u$, and $\bV_s=\bsigma_s\bsigma_s'$. If $\bJ_s$ has no pair of reverse-sign eigenvalues, then
\begin{align}
\vect{\bGamma_t}=(\bI\otimes\bJ_s+\bJ_s\otimes\bI)^{-1}\bv_t,\quad  \bv_t:=\exp\{\bJ_s(t-s)\}\bV_s\exp\{\bJ_s'(t-s)\}-\bV_s.\label{eq:gamma}
\end{align}
If $\bV_s^{-1}\bJ_s$ is symmetric, then $\bGamma_t$ admits a more
explicit form\footnote{Note the similar argument given in
  \cite{Roberts2002}, albeit in their equation (24) the covariance
  matrix is not symmetric, probably because of a typo in (25), which should have been
  $(J(x)a_{x,h})'=J(x)a_{x,h}$.}:
\begin{align}
\bGamma_t=&\,\int_s^t \exp\{\bJ_s(t-u)\}\bV_s
 \bV_s^{-1} \exp\{\bV_s\bJ_s'\bV_s^{-1}(t-u)\}\bV_s\rd u\nonumber\\
=&\,\frac{1}{2}\bJ_s^{-1}(\exp\{2\bJ_s(t-s)\}-\bI)\bV_s.\label{eq:closedgamma}
\end{align}
Interestingly, for the Langevin family of diffusions, $\bV_s^{-1}\bJ_s$
is guaranteed to be symmetric as long as the diffusion coefficient is
constant. This is due to the particular form of \eqref{eq:lange}, which
gives $\bJ_s = \frac{1}{2}\bV_s \bHcal_s$, where $\bHcal_s$ stands for
the Hessian of $\log f$ at $\bX_s$. Therefore, \eqref{eq:closedgamma}
simplifies notably the evaluation of the Shoji--Ozaki pseudo-likelihood
for all the toroidal diffusions considered in this paper. \\

The Shoji--Ozaki pseudo-tpd for toroidal diffusions \nolinebreak[4]is\footnote{In \cite{Shoji1998} the drift approximation is
done by It\^{o}'s formula. To obtain a simpler pseudo-likelihood, we use a local linear
approximation of $b$ as in \cite{Ozaki1985} (for the case $p=1$). Without this extra
simplification, the expectation becomes
$\tilde E_\Delta(\bphi)=E_\Delta(\bphi)+J(\bphi)^{-2}(\exp\{J(\bphi)\Delta\}-\bI-J(\bphi)\Delta)M(\bphi)$ with
$M(\bphi)=\frac{1}{2}\lrp{\tr{\bV(\bphi)\bH_1(\bphi)},\ldots,\tr{\bV(\bphi)\bH_n(\bphi)}}'$
and $\bH_i(\bphi)=\lrp{\tfrac{\partial^2b_i(\bphi)}{\partial \phi_k\partial\phi_l}}_{1\leq k,l\leq p}$,
$i=1,\ldots,p$.}
\[
p^\mathrm{SO}_{\Delta}(\btheta\,|\,\bphi)
:= f_\mathrm{WN}\lrp{\btheta;E_\Delta(\bphi),V_\Delta(\bphi)},\quad \btheta,\,\bphi\in\T^p,
\]
where, assuming that $V(\bphi)^{-1}J(\bphi)$ is symmetric (otherwise use
\eqref{eq:gamma} instead of \eqref{eq:closedgamma}),
\begin{align*}
E_\Delta(\bphi)=&\,\bphi+J(\bphi)^{-1}(\exp\{J(\bphi)\Delta\}-\bI)b(\bphi),\\ V_\Delta(\bphi)=&\,\frac{1}{2}J(\bphi)^{-1}(\exp\{2J(\bphi)\Delta\}-\bI)V(\bphi).
\end{align*}
When $J(\bphi)$ shrinks to $\zero$, then $E_\Delta(\bphi)\approx \bphi +
b(\bphi)\Delta$ and $V_\Delta(\bphi)\approx V(\bphi)\Delta$, so the Euler scheme follows by continuity. If all the real parts of the
eigenvalues of $J(\bphi)$ are negative, then
$
p^\mathrm{SO}_{\Delta}(\btheta\,|\,\bphi)\underset{\Delta\to\infty}{\longrightarrow}
f_\mathrm{WN}\big(\btheta;\bphi -
  J(\bphi)b(\bphi),-\frac{1}{2}J(\bphi)^{-1}V(\bphi)\big)
$
and the pseudo-tpd does not degenerate into a uniform
density as Euler's does. Otherwise, the pseudo-tpd converges to the uniform
distribution in $\T^p$ exponentially fast (see Figure \ref{fig:tpds}), at a rate controlled by the maximum positive real part of the eigenvalues.\\

A disadvantage of these pseudo-likelihoods is that
they are unimodal, so they cannot capture the multimodality of the
tpd, a distinctive feature of toroidal diffusions.

\begin{rem}
\label{rem:wn}
Evaluating $f_\mathrm{WN}(\cdot;\bmu,\bSigma)$ for the above pseudo-tpds is a computationally demanding task. Several approximations are possible:
\begin{enumerate}[label=\textit{\roman*}., ref=\textit{\roman*}]
\item \textbf{High-concentration}. Use the closest winding number as a one-term truncation of the series, \textit{i.e.}, \allowbreak$\phi_{\bSigma}(\cmod{\cdot-\bmu})$.\label{rem:wn:1}
\item \textbf{Fixed truncation}. \citet[page 50]{Mardia2000} suggests (for $p=1$) $\sum_{\bk\in\{-1,0,1\}^p}\phi_{\bSigma}(\cdot-\bmu+2\bk\pi)$, which is usually enough for practical purposes if the argument lays in $\T^p$.\label{rem:wn:2}
\item \textbf{Von Mises moment matching}. Uses the approximation $\mathrm{WN}(\mu,\sigma^2)\approx \mathrm{vM}(\mu,A_1^{-1}(e^{-\sigma^2/2}))$, with \allowbreak$A_1(\kappa)=\Ical_1(\kappa)/\Ical_0(\kappa)$ \citep[page 38]{Mardia2000}. This approximation generalizes easily to the multivariate case only if $\bSigma$ is diagonal. For the bivariate case, an alternative is to use a von Mises \emph{score} matching \citep{Mardia2017}.\label{rem:wn:3}
\item \textbf{Adaptive truncation}. The \cite{Jona-Lasinio2012}'s ``$3\sigma$ adaptive truncation'' can be generalized to the multivariate case by Bonferroni: $\sum_{\bk=\bk_L}^{\bk_U}\phi_{\bSigma}(\cdot-\bmu+2\bk\pi)$ with $\bk_U=-\bk_L=1+\lfloor z_{1-\alpha/(2p)}\sqrt{\diag{\bSigma}}\allowbreak/(2\pi)\rfloor$, where $z_{\alpha}$ is the upper $\alpha$-quantile of a $\mathcal{N}(0,1)$, ensures a probability mass in $\T^p$ larger than $1-\alpha$.\label{rem:wn:4}

\end{enumerate}
For $p=1,2$, a simple compromise between tractability and accuracy is combining \ref{rem:wn:1} and \ref{rem:wn:2} into $\sum_{\bk\in\{-1,0,1\}^p}\allowbreak\phi_{\bSigma}(\cmod{\cdot-\bmu}+2\bk\pi)$, which has a probability coverage of $\T^p$ larger than\linebreak $1-2\sum_{j=1}^p\Phi(-\tfrac{3\pi}{\sigma_j})$.
\end{rem}

%-------------------------------------%
\subsection{Wrapped Ornstein--Uhlenbeck approximation of the WN process}
%-------------------------------------%

We now present a specific approximation for the tpd of the WN process that allows to model the multimodality in the tpd. Multimodality is not uncommon for toroidal diffusions since each coordinate can move towards its mean value in two directions and, contrary to what happens with the OU process, this implies that neither the WN nor the MvM processes have tpds within the parametric\nopagebreak[4] families of the sdis.\\

The approximation relies on the connection of the WN process with the tractable multivariate OU process:
\begin{align}
\rd\bX_t=\bA(\bmu-\bX_t)\rd t+\bSigma^\frac{1}{2}\rd \bW_t,\label{eq:mou}
\end{align}
with $\bmu\in\R^p$, $\bSigma$ a covariance matrix, and $\bA$ such that $\bA^{-1}\bSigma$ is a covariance matrix. The last
  assumption ensures that the OU process is ergodic and time-reversible,
and as a consequence, implies a simple expression for the
  covariance matrix of the tpd (see below). Under this setting, the process is ergodic, time-reversible, and has stationary density $\mathcal{N}\big(\bmu,\frac{1}{2}\bA^{-1}\bSigma\big)$. We denote by WOU, standing for Wrapped multivariate OU process, to the wrapping of \eqref{eq:mou}. Assuming that $\bX_s\sim \mathcal{N}\big(\bmu,\frac{1}{2}\bA^{-1}\bSigma\big)$, the conditional density of WOU is given by Proposition \ref{prop:condden} and the tpd of \eqref{eq:mou}:
\begin{align}
p^\mathrm{WOU}_t(\btheta\,|\,\btheta_s) := \sum_{\bm\in\Z^p} f_\mathrm{WN}(\btheta;\bmu^\bm_t,\bGamma_t)w_{\bm}(\btheta_s), \quad w_{\bm}(\bx)=\frac{\phi_{\frac{1}{2}\bA^{-1}\bSigma}(\bx-\bmu+2\bm\pi )}{\sum_{\bk\in\Z^p}
  \phi_{\frac{1}{2}\bA^{-1}\bSigma}(\bx-\bmu+2\bk\pi)}\label{eq:wou}
\end{align}
where, by the same argument used in \eqref{eq:closedgamma},
\begin{align*}
\bmu^\bm_t := \bmu + e^{-t\bA }(\btheta_s - \bmu + 2\bm\pi ),\quad \bGamma_t=\int_0^te^{-s\bA}\bSigma e^{-s\bA'}\rd s=\frac{1}{2}\bA^{-1}(\bI-\exp\{-2t\bA\})\bSigma.
\end{align*}
The conditional density \eqref{eq:wou} can be seen as a wrapping of the tpd of \eqref{eq:mou} weighted by the sdi of the winding numbers, which resembles the structure of the WN process drift: a weighting of linear drifts like \eqref{eq:mou} according to the winding number sdi in order to achieve periodicity. Albeit \eqref{eq:wou} and the tpd of the WN process are different, they behave similarly in many situations. The next corollary from Proposition \ref{prop:condden} formalizes these arguments.
\begin{coro}
\label{coro:wou}
{\em Suppose $\{\bTheta_t\}$ solves \eqref{eq:wnp} with $\bTheta_0 =
  \btheta_0$ and let $\{\bTheta^\mathrm{WOU}_t\}$ be the wrapping of the
solution to \eqref{eq:mou}, where $\bX_0\sim\mathcal{N}(\bmu,\frac{1}{2}\bA^{-1}\bSigma)$. We condition, moreover, on $\bTheta^\mathrm{WOU}_0 =
\btheta_0$. Then:
\begin{enumerate}[label=\textit{\roman*.}, ref=\textit{\roman*}]
\item As $t \rightarrow 0$, $\bTheta_t\rightarrow\btheta_0$ and $\bTheta^\mathrm{WOU}_t\rightarrow\btheta_0$ in probability.
\item As $t \rightarrow \infty$, both $\bTheta_t$ and $\bTheta^\mathrm{WOU}_t$
  converge in distribution to a
  $\mathrm{WN}(\bmu,\tfrac{1}{2}\bA^{-1}\bSigma)$.
\item When $\bA^{-1}\bSigma \rightarrow \mathbf{0}$ with $\bSigma$ bounded,
$\bTheta_t - \bTheta^\mathrm{WOU}_t \rightarrow \mathbf{0}$ in probability, so the
distributions of $\bTheta_t$ and $\bTheta^\mathrm{WOU}_t$ are similar in the limit.
\item $p_t^\mathrm{WOU}$ satisfies
$f_{\mathrm{WN}}(\btheta_0;\bmu,\tfrac{1}{2}\bA^{-1}\bSigma)
p^\mathrm{WOU}_t(\btheta\,|\,\btheta_0) =
f_{\mathrm{WN}}(\btheta;\bmu,\tfrac{1}{2}\bA^{-1}\bSigma)
p^\mathrm{WOU}_t(\btheta_0\,|\,\btheta)$,
$\forall \btheta,\btheta_0\in\T^p$ (just like $p_t$).
\end{enumerate}}
\end{coro}
\begin{proof}
The first two statements for the WN process are well-known for any diffusion, and it
follows from \eqref{eq:wou} that, for $\btheta
\neq \btheta_0$, $\lim_{t\to 0}p^\mathrm{WOU}_t(\btheta\,|\,\btheta_0) = 0$ and that $\lim_{t\to\infty}
p^\mathrm{WOU}_t(\btheta\,|\,\btheta_0) =
f_{\mathrm{WN}}(\btheta;\bmu,\frac{1}{2}\bA^{-1}\bSigma)$. The last
statement follows from \eqref{eq:wou} and the fact that the OU process is time-reversible when
$\bA^{-1}\bSigma$ is positive definite. We give a rough
sketch of a proof of the third statement. The result follows because
the tpd of the WN process is asymptotically equal to the tpd of the OU
process in the high concentration limit. To see this, suppose that
$\bX_t$ solves \eqref{eq:mou} with $\bX_0 = \btheta_0$ (we can ignore
the other starting points), and that $\bTheta_t$ solves \eqref{eq:wnp}
with $\bTheta_0 = \btheta_0$, both driven by the same Wiener
process. Then $\bY_t:= \bX_t - \bTheta_t$ solves $\rd\bY_t = -\bA
\bY_t \rd t + \rd \bZ_t$, with $\bY_0 = \zero$, where
\begin{align*}
\bZ_t =&\, -\bA \sum_{\bk\in\Z^p\backslash \{\zero\}}\int_0^t \lrp{\bmu-\bTheta_s-2\bk\pi}
w_{\bk}(\bTheta_s) \rd s + \bA \int_0^t \lrp{\bmu-\bTheta_s}(1-w_{\zero}(\bTheta_s)) \rd s.
\end{align*}
If $\bY_t = \int_0^t e^{-\bA(t-s)}\rd\bZ_s \rightarrow \zero$ in probability, then the two distributions of $\bTheta_t$ and $\bX_t$ will be the same in the limit. This follows because
$\bA w_{\bk}(\btheta) \rightarrow \zero$ and $\bA (1- w_{\zero}(\btheta)) \rightarrow
\zero$ for $\bk \in \Z^p\backslash \{ \zero\}$ and $-\pi < \btheta < \pi$ (we consider only the case $\mathrm{wind}(\bmu-\btheta) = \zero$ because $\mathbb{P}[|\bTheta_s-\bmu|\leq \boldsymbol{\pi}]\to1$), and hence $\bZ_s \rightarrow \zero$ in probability for all $s \leq t$.
\end{proof}

The tractability of \eqref{eq:wou} degenerates quickly with the dimension, but it can be readily computed for $p=1,2$, two highly relevant situations in practice. We focus our attention on implementation matters for the non-trivial case $p=2$. The first point of inquiry is what parametrization of $\bA$ and $\bSigma$ leads to a covariance matrix $\bA^{-1}\bSigma$, which guarantees a non-degenerate WN sdi.
\begin{lem}
Let $\bA$ and $\bSigma$ be $2\times 2$ matrices, $\bSigma=\big(
\sigma_1^2,\allowbreak \rho\sigma_{1}\sigma_{2};
\rho\sigma_{1}\sigma_{2}, \sigma_2^2\big)$ positive-definite. Assume $\alpha_1,\alpha_2>0$. Any matrix $\bA$ such that $\bA^{-1}\bSigma$ is a covariance matrix has the form
\[
\bA=\begin{pmatrix}
\alpha_1 & \frac{\sigma_1}{\sigma_2}\lrp{\alpha_3+\tfrac{1}{2}\rho(\alpha_2-\alpha_1)}\\
\frac{\sigma_2}{\sigma_1}\lrp{\alpha_3-\tfrac{1}{2}\rho(\alpha_2-\alpha_1)} & \alpha_2\\
\end{pmatrix},
\]
with $\alpha_3^2<\frac{\rho^2(\alpha_1-\alpha_2)^2}{4}+\alpha_1\alpha_2$.
\end{lem}
The parametrization with $\rho=0$ provides a compromise between flexibility and simplicity, and will be employed throughout (first occurrences in Figures \ref{fig:tpds} and \ref{fig:klwn2d}). With $\rho=0$ the dependence between components is modelled by $\alpha_3$, which is clear from
\[
\frac{1}{2}\bA^{-1}\bSigma=\frac{1}{2(\alpha_1\alpha_2-\alpha_3^2)}
\begin{pmatrix}
\alpha_2 \sigma_1^2 & -\alpha_3 \sigma_1 \sigma_2 \\
-\alpha_3 \sigma_1 \sigma_2 & \alpha_1 \sigma_2^2 \\
\end{pmatrix}.
\]

The second point is the efficient computation of $e^{-t\bA}$ and $\bGamma_t$. In virtue of Corollary 2.4 of \cite{Bernstein1993}, $e^{t\bA}=a(t)\bI + b(t)\bA$ with $b(t):=e^{s(\bA)t}\tfrac{\sinh(q(\bA)t)}{q(\bA)}$ (if $q(\bA)=0$, then, by continuity, $b(t)=e^{s(\bA)t}t$), $a(t):=e^{s(\bA)t}\cosh(q(\bA)t)-s(\bA)b(t)$, $s(\bA):=\tfrac{\tr{\bA}}{2}$, and $q(\bA):=\sqrt{\abs{\det(\bA-s(\bA)\bI)}}$. Therefore,
\begin{align}
\bGamma_t
=&\,s(t)\frac{1}{2}\bA^{-1}\bSigma + i(t)\bSigma,\label{eq:closedgamma2}
\end{align}
with $s(t):=1-a(-2t)$ and $i(t):=-\tfrac{1}{2}b(-2t)$. Expression \eqref{eq:closedgamma2} shows neatly the interpolation between the infinitesimal and stationary covariance matrices and is especially useful if it is required to compute the tpd for several $t$'s.\\

To conclude, we highlight some of the advantages of the WOU approximation over the Euler and Shoji--Ozaki pseudo-likelihoods for the WN process. Firstly, WOU is able to capture the multimodality of the tpd (see Figure \ref{fig:tpds}) and has the correct sdi. Secondly, WOU is faster to compute than Shoji--Ozaki, as it does not require exponentiation and inversion of the Jacobian matrix for each observation, but only once.

%-------------------------------------%
\subsection{Likelihood by numerical PDE solution}
\label{sec:est:pde}
%-------------------------------------%

An alternative to approximate likelihoods is to compute the ``exact'' (up to a prescribed accuracy) MLE by a numerical solution of \eqref{eq:pde}. This approach is computationally expensive, but remains valid for arbitrary diffusions and discretization times. Moreover, it provides insightful visualizations of the tpd. In the following, we discuss how to solve numerically \eqref{eq:pde} for dimensions $p=1,2$.

%-------------------------------------%
\subsubsection{One-dimensional case}
%-------------------------------------%

We consider a state grid $\mathcal{G}:=\lrb{x_1,\ldots,x_{M_x}}$ in $[-\pi,\pi)$ constructed with step $\Delta x:=\frac{2\pi}{M_x}$, and such that $x_{M_x+1}\allowbreak:=x_{1}=-\pi$ and $x_{0}:=x_{M_x}=\pi-\Delta x$. We also consider a time grid in $[0, T]$ with $\Delta t:=\frac{T}{M_t}$. For consistency with the common notation for PDEs, we refer by $u(x,t)$ to $p_t(x\,|\,p_s):=\int_{\T^1}p_t(x\,|\,\phi)p_s(\phi)\rd\phi$, the solution of the PDE for the initial condition (at time $s$) given by a circular density $p_s$. The vector $\bu^n$, $n=0,\ldots,{M_t}$, denotes the tpd evaluated at $\mathcal{G}$ at time $s+n\Delta t$. We write $b_{i}:=b(x_i)$ and $\sigma^2_i:=\sigma^2(x_i)$, $i=1,\ldots,{M_x}$.\\

We employ the so-called \textit{Crank--Nicolson scheme} for discretizing \eqref{eq:pde}, which can be rewritten as
\begin{align*}
\pf{u}{t}=-\pf{}{x}(b(x)u(x,t))+\frac{1}{2}\pftwo{}{x}(\sigma^2(x)u(x,t))=F\lrp{u,x,t,\pf{u}{x},\pftwo{u}{x}}.
\end{align*}
Crank--Nicolson is a well-known scheme for diffusion and convection-diffusion PDEs such as \eqref{eq:pde}. It is based on a trapezoidal-like approximation of the forward difference of the time derivative that is combined with a centered finite differences of the state derivatives:
\begin{align*}
\frac{u_i^{n+1}-u_i^{n}}{\Delta t}=\frac{1}{2}\lrc{F_i^{n+1}+F_i^{n}},\quad F_i^{n}:=\frac{2}{\Delta t}\lrp{\gamma_i u_{i+1}^{n}-2\beta_i u_{i}^{n}+\alpha_i u_{i-1}^{n}},
\end{align*}
with $r:=\frac{\Delta t}{4(\Delta x)^2}$, $\gamma_i:=\lrp{-b_{i+1}\Delta x + \sigma^2_{i+1}}r$, $\beta_i:=\sigma_i^2r$, and $\alpha_i:=\lrp{b_{i-1}\Delta x + \sigma^2_{i-1}}r$. The next step in time of the solution, $\bu^{n+1}$, is implicitly given by the system
\begin{align}
(\mathbf{F}-\bI)&\bu^{n+1}=\mathbf{d}^n,\quad \mathbf{F}:=\begin{pmatrix}
-2\beta_1 & \gamma_1                     &         & \alpha_1\\
\alpha_2      & -2\beta_2       &    \ddots    & \\
                            & \ddots        & \ddots & \gamma_{{M_x}-1}\\
\gamma_{M_x} &   & \alpha_{M_x} & -2\beta_{M_x}
\end{pmatrix},\nonumber\\
\mathbf{d}^n:=&\,-(\mathbf{F}+\bI)\bu^n =-\boldsymbol\gamma\circ\bu_+^{n}+(2\boldsymbol\beta-1)\circ\bu^{n}-\ba\circ\bu_-^{n},\label{eq:trigdiag}
\end{align}
with subscript $\pm$ denoting the vector with entries circularly shifted $\mp1$ position. It is well-known \citep[page 225]{Thomas1995} that this periodic tridiagonal system can be solved efficiently by tacking the tridiagonal systems $\bB\by_1=\mathbf{d}_n$ and $\bB\by_2=\mathbf{w}$ (where $\bF=\bB - \mathbf{w}\bz'$), and using the Sherman--Morrison formula: $\bu^{n+1}=\by_1+\tfrac{\bz'\by_1}{1-\bz'\by_2}\by_2$. The latter tridiagonal systems can be jointly solved by a modification of the Thomas algorithm, since they share coefficient matrix. The cost of the solution is $\Order{{M_t}{M_x}}$.
In addition, since $\mathbf{F}$ is constant with respect to time, the \textit{tridiagonal} LU factorization underlying the Thomas algorithm can be reused, yielding a complexity factor reduction of $5/8$ on the tridiagonal solver.

%-------------------------------------%
\subsubsection{Two-dimensional case}
%-------------------------------------%

We consider now two grids $\mathcal{G}_x$ and $\mathcal{G}_y$ analogous to $\mathcal{G}$, but of sizes $M_x$ and $M_y$, and steps $\Delta x$ and $\Delta y$. We refer by $u(x,y,t)$ to $p_t(x,y\,|\,p_s):=\int_{\T^2}p_t(x,y\,|\,\bphi)p_s(\bphi)\rd\bphi$, where $p_s$ is a toroidal density giving the initial condition (at time $s$). The matrix $\bU^n$,
$n=0,\ldots,{M_t}$, denotes the tpd evaluated at $\mathcal{G}_x\times\mathcal{G}_y$ at time $s+n\Delta t$. We write $b_{z;i,j}:=b_z(x_i,y_j)$, $\sigma_{z;i,j}^2:=\sigma_z^2(x_i,y_j)$, $\sigma_{xy;i,j}^2:=\sigma_{xy}^2(x_i,y_j)$, with $z$ standing for $x$ or $y$, and $i=1,\ldots,M_x$, $j=1,\ldots,M_y$. With this notation, \eqref{eq:pde} becomes
\begin{align*}
\pf{u}{t}=&\,\sum_{z\in\{x,y\}}\Big[-\pf{}{z}(b_z(x,y)u(x,y,t))+\frac{1}{2}\pftwo{}{z}(\sigma_z^2(x,y)u(x,y,t))\Big]+\pfmix{}{x}{y}(\sigma_{xy}(x,y)u(x,y,t))
\end{align*}
The Crank--Nicolson scheme proceeds as in the one-dimensional case:
\begin{align}
\frac{u_{i,j}^{n+1}-u_{i,j}^{n}}{\Delta t}=\frac{1}{2}\lrc{F_{i,j}^{n+1}+F_{i,j}^{n}},\label{eq:us}
\end{align}
with finite differences that can be collected into three terms associated to the partial and mixed derivatives:
\begin{align*}
F_{i,j}^{n}:=&\,F^n_{x;i,j}+F^n_{y;i,j}+F^n_{xy;i,j},\\
F^n_{x;i,j}:=&\,\frac{2}{\Delta t}\lrp{\Gamma_{x;i,j} u_{i+1,j}^{n}-2B_{x;i,j} u_{i,j}^{n}+A_{x;i,j} u_{i-1,j}^{n}},\\
F^n_{y;i,j}:=&\,\frac{2}{\Delta t}\lrp{\Gamma_{y;i,j} u_{i,j+1}^{n}-2B_{y;i,j} u_{i,j}^{n}+A_{y;i,j} u_{i,j-1}^{n}},\\
F^n_{xy;i,j}:=&\,\frac{2}{\Delta t}\big(C^{+,+}_{i,j} u_{i+1,j+1}^{n}-C^{+,-}_{i,j} u_{i+1,j-1}^{n} -C^{-,+}_{i,j} u_{i-1,j+1}^{n}+C^{-,-}_{i,j} u_{i-1,j-1}^{n}\big).
\end{align*}
We have denoted $r_z:=\frac{\Delta t}{4(\Delta z)^2}$, $r_{xy}:=\frac{\Delta t}{8\Delta x\Delta y}$, and
\begin{align*}
\Gamma_{x;i,j}:=&\,(-b_{x;i+1,j}\Delta x+\sigma^2_{x;i+1,j})r_x,\\
\Gamma_{y;i,j}:=&\,(-b_{y;i,j+1}\Delta y+\sigma^2_{y;i,j+1})r_y,\\
A_{x;i,j}:=&\,(b_{x;i-1,j}\Delta z+\sigma^2_{x;i-1,j})r_x,\\
A_{y;i,j}:=&\,(b_{y;i,j-1}\Delta y+\sigma^2_{y;i,j-1})r_y,\\
B_{z;i,j}:=&\,\sigma^2_{z;i,j}r_z,\quad C^{\pm,\pm}_{i,j}:=\sigma_{xy;i\pm1,j\pm1}r_{xy}.
\end{align*}
Let $F=F_x+F_y+F_{xy}$ be the linear functions mapping $\bU^n$ into $F(\bU^n)=F_x(\bU^n)+F_y(\bU^n)+F_{xy}(\bU^n)=(F^n_{x;i,j}+F^n_{y;i,j}+F^n_{xy;i,j})=(F^n_{i,j})$ and $I$ the identity function. Then, we can express \eqref{eq:us}\nolinebreak[4] as
\begin{align}
(F_x+F_y+F_{xy}-I)(\bU^{n+1})=-(F+I)(\bU^{n}).\label{eq:FU}
\end{align}
If the left and right hand sides of \eqref{eq:FU} are stacked column-wise, \eqref{eq:FU} becomes an $M_xM_y\times M_xM_y$ periodic $9$-diagonal system. This system cannot be solved so efficiently as in the tridiagonal case, requiring a more complex algorithm or a generic sparse LU factorization. \\

An alternative approach that reduces drastically the computational burden of solving \eqref{eq:FU} is to adopt an Alternating Direction Implicit (ADI) scheme. ADI schemes split the multidimensional finite differences in a series of univariate discretizations with simpler associated systems. Originally developed for the diffusion equation, they were extended to the convection-diffusion equations with a mixed de\-riv\-ative term by \cite{McKee1996}, in the so-called \textit{Douglas scheme}. This scheme proceeds with an explicit multivariate step corrected by two unidimensional Crank--Nicolson steps, whose purpose is to stabilize the explicit step:
\begin{align}
\bY_0&=\bU^{n}+\Delta t F(\bU^{n})\quad\text{ (explicit)}\label{eq:doug1}\\
\bY_1-\tfrac{\Delta t}{2} F_x(\bY_1)&=\bY_0-\tfrac{\Delta t}{2}F_x(\bU^{n})\quad\text{(implicit)}\label{eq:doug2}\\
\bY_2-\tfrac{\Delta t}{2} F_y(\bY_2)&=\bY_1-\tfrac{\Delta t}{2}F_y(\bU^{n})\quad\text{(implicit)}\label{eq:doug3}\\
\bU^{n+1}&=\bY_2\nonumber
\end{align}
Consequently, if the matrix equations in \eqref{eq:doug1}--\eqref{eq:doug3} are transformed into linear systems by column-wise stacking for \eqref{eq:doug1} and \eqref{eq:doug2}, and row-wise stacking for \eqref{eq:doug3}, the Douglas scheme transforms the difficult task of solving \eqref{eq:FU} into solving two periodic tridiagonal systems of size $M_xM_y$. Specifically, the steps in \eqref{eq:doug1}--\eqref{eq:doug3} are carried out using
\begin{align*}
\mathrm{vec}(F_{xy}(\bY))=&\,\mathrm{vec}(\mathbf{C}^{+,+})\circ\by_{+,+,c}-\mathrm{vec}(\mathbf{C}^{+,-})\circ\by_{+,-,c}\\
&-\mathrm{vec}(\mathbf{C}^{-,+})\circ\by_{-,+,c}+\mathrm{vec}(\mathbf{C}^{-,-})\circ\by_{-,-,c},\\
\by_{\pm,\pm,c}:=&\,\mathrm{vec}(\bY_{\pm,\pm})\\
\mathrm{vec}(F_x(\bY))=&\,\mathrm{vec}(\bGamma_x)\circ\by_{c,+}-2\mathrm{vec}(\bB_x)\circ\by_c+\mathrm{vec}(\bA_x)\circ\by_{c,-},\\
\mathrm{vec}(F_y(\bY)')=&\,\mathrm{vec}(\bGamma_y')\circ\by_{r,+}-2\mathrm{vec}(\bB_y')\circ\by_r+\mathrm{vec}(\bA_y')\circ\by_{r,-},\\
\by_c:=&\,\mathrm{vec}(\bY),\quad \by_r:=\mathrm{vec}(\bY').
\end{align*}
$\bU^{n+1}$ is obtained by setting $\bY$ equal to $\bU^n$, $\bY_1$ or $\bY_2$ in the above expressions and by solving \eqref{eq:doug2} and \eqref{eq:doug3} as \eqref{eq:trigdiag} was. Then, the total cost of the solution is $\Order{{M_t}M_xM_y}$. Note that the row-stacking vector $\by_r^n$ can be directly obtained from $\by_c^n$ by extracting the indexes $((k_c-1)\mod M_y)M_x+\big\lfloor\frac{k_c - 1}{M_y}\big\rfloor+1$, $k_c=1,\ldots,\allowbreak M_xM_y$, of the latter (analogous for the converse). We refer to the neat expository paper of \cite{IntHout2010} for further description of ADI schemes.

%-------------------------------------%
\subsubsection{Remarks on the discretization schemes}
%-------------------------------------%

The Crank--Nicolson and Douglas schemes are tailored solutions for solving \eqref{eq:pde} that exploit the particular PDE structure. It is worth to note that, among other methods, a well-known approach to solve PDEs is the \textit{method of lines}. This method is prone to create stiff systems, which need to be handled properly by a meta-solver that chooses between stiff and non-stiff solvers (e.g., the \texttt{lsoda} implementation in \cite{Soetaert2012}). Not surprisingly, in our application we found that the efficiency and reliability of the tailored solutions were superior to the latter, much more general, meta-solver. \\

Some theoretical remarks about the schemes employed are given as follows. The Crank--Nicolson scheme is \textit{conservative} (hence the Douglas scheme is too), which can be easily seen from the periodic tridiagonal system. It is also second-order consistent in time and space (with the discretization used), with the appealing property of being \textit{unconditionally stable} with respect to $\Delta t$. The Douglas scheme is first-order consistent and unconditionally stable when applied to two-dimensional convection-diffusion equations with a mixed derivative term. See \cite{IntHout2010} for the description of second-order ADI schemes of the same computational complexity (but with a factor increase of at least two). Both unconditional stabilities refer to the usual framework of constant coefficients.

\begin{figure}[H]
\centering
\includegraphics[width=\textwidth]{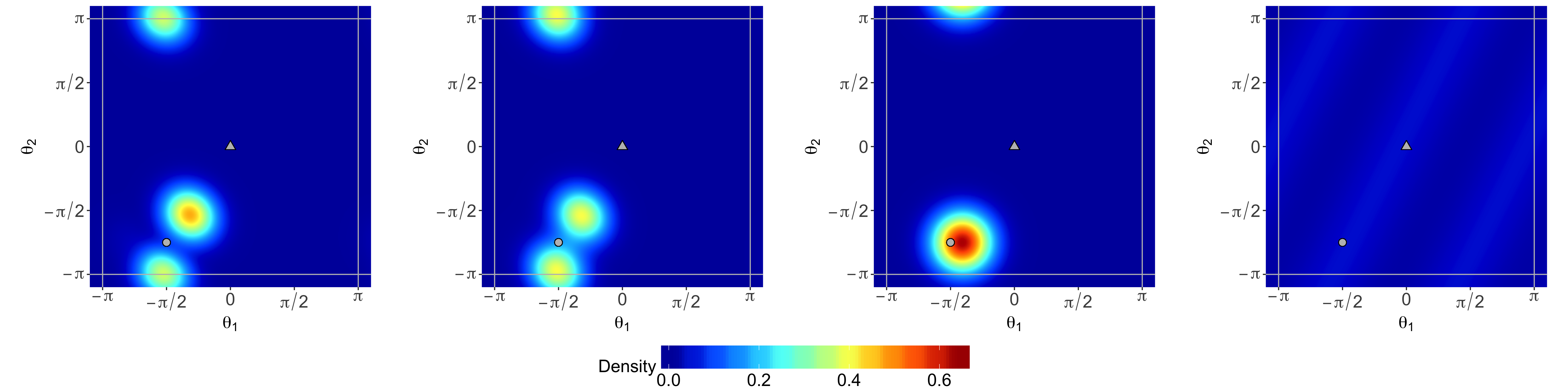}
\caption{\small The four approximations to the tpd $p_t(\cdot\,|\,\btheta_0)$ for the two-dimensional WN process, with $\btheta_0=(-\frac{\pi}{2}, -\frac{3\pi}{4})$ (round facet), and $t=0.25$. From left to right: PDE solution with $\sigma_0=0.01$, WOU tpd approximation, Euler pseudo-tpd, and Shoji--Ozaki pseudo-tpd. The WN process has parameters $\ba=(1,1,0.5)$,  $\bmu=(0,0)$ (triangular facet), and $\bSigma=\diag{1, 1}$.\label{fig:tpds}}
\end{figure}

%-------------------------------------%
\subsubsection{Likelihood evaluation}
%-------------------------------------%

The PDE numerical solutions approximate $p_t(\btheta\,|\,p_s)=\int_{\T^p}p_t(\btheta\,|\,\bphi)\allowbreak p_s(\bphi)\rd\bphi$, where $p_s$ is a density over $\T^p$ giving the initial condition. Therefore, $p_t(\btheta\,|\,\btheta_0)$ can be approximated by considering a concentrated $\mathrm{WN}(\btheta_0,\sigma^2_0\bI)$ as the initial condition. For a fixed grid, $\sigma_0$ must not be set to an arbitrarily small value, as it will create a sharp initial condition poorly discretized and prone to raise numerical errors. A possible rule of thumb is to choose a small $\sigma_0$ such that the periodic trapezoidal rule of the discretized  $\mathrm{WN}(\btheta_0,\sigma^2_0\bI)$ is close to one. \\

We illustrate the evaluation of the log-likelihood \eqref{eq:ll} from the PDE solution for $p=1$. The extension to $p=2$ is conceptually straightforward, albeit cumbersome in notation. Given the sample $\{\Theta_{\Delta i}\}_{i=1}^N$ and the grid $\mathcal{G}$, let denote by $\mathbf{P}:=p_t(\mathcal{G}\,|\,\mathcal{G})$ the $M_x\times M_x$ \textit{tpd matrix} of the process discretized in $\mathcal{G}$. The $j$-th column of $\mathbf{P}$ is obtained by solving the PDE with initial condition $\mathrm{WN}(x_j,\sigma^2_0)$. We can approximate $p_\Delta(\Theta_{i\Delta}\,|\,\Theta_{(i-1)\Delta})$ from $\mathbf{P}$ by linear interpolation:
\begin{align}
p_\Delta&(\Theta_{i\Delta}\,|\,\Theta_{(i-1)\Delta}) \approx\sum_{k,l=0}^1 \omega_{k}(\Theta_{\Delta i})P_{g_0(i)+k,g_0(i-1)+l}\omega_{l}(\Theta_{\Delta (i-1)}),\label{eq:linint}
\end{align}
with $g_0(i):=\lceil\frac{\Theta_{\Delta i}+\pi}{\Delta x}\rceil$, $\omega_0(\theta)=\frac{x_{g_0(i)+1}-\theta}{\Delta x}$, and $\omega_1(\theta)=\frac{\theta-x_{g_0(i)}}{\Delta x}$. The log-likelihood is obtained by plugging \eqref{eq:linint} into \eqref{eq:ll}. The advantage of doing so is that the number of PDE solutions required for a single log-likelihood evaluation remains bounded by $M_x$, irrespectively of $N$. In addition, we only need to compute the columns of $\mathbf{P}$ corresponding to the unique set of indexes $\{g_0(i)+l:i=0,\ldots,N-1,l=0,1\}$. A simpler, though less precise, alternative to \eqref{eq:linint} is to use constant interpolation for $\Theta_{\Delta (i-1)}$. This results in a lower number of PDE solutions, specially in the two-dimensional case. Finally, if the drift is antisymmetric around a point $\mu$, then $p_t(\theta\,|\,\varphi)=p_t(2\mu-\theta\,|\,2\mu-\varphi)$. Hence, if $\mathcal{G}$ is circularly centered at $\mu$, half of the columns of $\mathbf{P}$ contain redundant information. The situation is analogous for $p=2$: if $b(\theta_1-\mu_1,\theta_2-\mu_2)=-b(\mu_1-\theta_1,\mu_2-\theta_2)$, $\forall\theta_1,\theta_2\in[-\pi,\pi)$, and $\mathcal{G}_x$ and $\mathcal{G}_y$ are both centered at $\mu_1$ and $\mu_2$, respectively, then only half of the columns of $\mathbf{P}$ are required. If the drift is isotropic, then only one fourth of the columns are needed.

%-------------------------------------%
\section{Simulation study}
\label{sec:sim}
%-------------------------------------%

We measure now the performance of the likelihood approximations given in Section \ref{sec:est}. Two types of empirical analysis are employed. First, we compare the divergences between the true tpd of a diffusion and its approximations across time. Second, we examine the errors of the approximate likelihoods in estimating $\blambda$ in several diffusions.

%-------------------------------------%
\subsection{Kullback--Leibler divergences for WN and vM processes}
%-------------------------------------%

All the estimation approaches described on Section \ref{sec:est} share a common root: the substitution of the true tpd $p_t$ by an approximation $p^\mathrm{A}_t$. The goodness-of-fit of these approximations has a direct influence on MLE since, for a general parametric setting, MLE is equivalent to minimizing the Kullback--Leibler divergence of the parametric pdf from the empirical pdf. We propose to measure the Kullback--Leibler divergence of $p^\mathrm{A}_t(\cdot\,|\,\btheta_s)$ from $p_t(\cdot\,|\,\btheta_s)$ by \textit{weighting} with the stationary density the contributions of each initial point $\btheta_s$:
\[
\mathrm{D}^\mathrm{A}_t=\int_{\T^p}\int_{\T^p}p_t(\btheta\,|\,\btheta_s)\log\lrp{\frac{p_t(\btheta\,|\,\btheta_s)}{p^\mathrm{A}_t(\btheta\,|\,\btheta_s)}}\nu(\btheta_s)\rd\btheta\rd\btheta_s.
\]

The curve $\mathrm{D}^\mathrm{A}_t$ gives a succinct summary of the goodness-of-fit of any approximation to the tpd across time. Its effective computation -- when no analytical expression for the tpd exists -- can be done with the PDE solution to the tpd. Some care is needed though. The PDE solution involves the initial condition in the form of a concentrated $\mathrm{WN}(\btheta_0,\sigma_0^2\bI)$. This initial condition implies that the PDE solution is approximating $p_{t,\sigma_0^2}(\bthe\,|\,\btheta_0):=\int_{\T^p}p_t(\bthe\,|\,\bphi)f_\mathrm{WN}(\bphi;\bthe_0,\sigma_0^2)\rd \bphi$ rather than $p_t$. Therefore, a more adequate approach is to smooth also the approximations in the computation of $\mathrm{D}^\mathrm{A}_t$ to perform a fair comparison:
\begin{align*}
\mathrm{D}^\mathrm{A}_{t,\sigma_0^2}=\int_{\T^p}\int_{\T^p}&u_{t,\sigma_0^2}(\btheta\,|\,\btheta_s)\log\lrp{\frac{u_{t,\sigma_0^2}(\btheta\,|\,\btheta_s)}{p^\mathrm{A}_{t,\sigma_0^2}(\btheta\,|\,\btheta_s)}} \nu(\btheta_s)\rd\btheta\rd\btheta_s.
\end{align*}

We explore the $\mathrm{D}^\mathrm{A}_{t,\sigma_0^2}$ curves for several variants of the approximations given in Section \ref{sec:est}, denoted as S (Stationary density), E (Euler), SO (Shoji--Ozaki), UE (Unwrapped Euler -- the usual Euler pseudo-likelihood), USO (Unwrapped Shoji--Ozaki), EvM, SOvM, and WOU. Suffix vM denotes the use of a vM distribution (\textit{moment} if one-dimensional; \textit{score} if two-dimensional) \textit{matching} approximation to the WN distribution appearing in the pseudo-likelihoods. 

\begin{figure}[!h]
	\vspace*{-0.25cm}
	\centering
	\includegraphics[width=.90\textwidth]{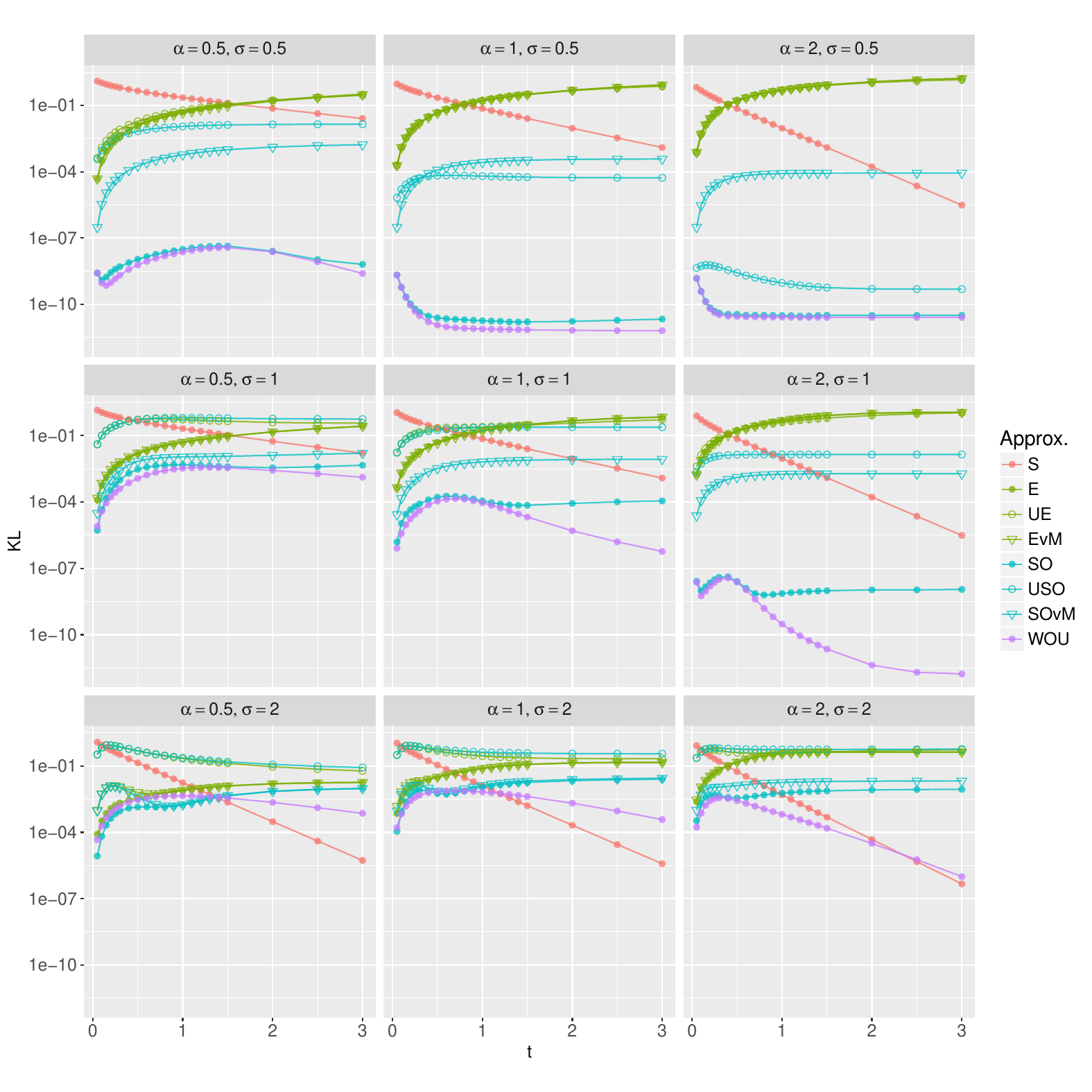}
	\vspace*{-0.5cm}
	\caption{\small $\mathrm{D}^\mathrm{A}_{t,\sigma_0^2}$ curves for $p=1$ on the WN process, on vertical log-scale. From top to bottom, rows represent small, moderate, and high diffusivities. The PDE was solved with $M_x=3000$, $M_t=\lceil 1500t\rceil$, and $\sigma_0=0.1$.}
	\label{fig:klwn1d}
\end{figure}

Figures \ref{fig:klwn1d} and \ref{fig:klwn2d} show the Kullback--Leibler curves for the WN process with $p=1$ and $p=2$, under different drift strengths and diffusivities. We highlight as follows their main features. First, WOU outperforms in almost all scenarios and times the other approximations. The main exceptions are the lower left scenarios of both figures, representing processes with a high diffusivity (small drifts and large diffusivities), where WOU is outperformed by SO and E for a significant range of intermediate times. In addition to S, WOU is the only approximation whose accuracy improves as time increases, above a certain local maximum in the Kullback--Leibler divergence. Second, the Euler and Shoji--Ozaki pseudo-likelihoods deteriorate or stabilize as time increases, except for scenarios with low and moderate diffusivity where SO is close to WOU (and both are close to the true tpd). E is systematically behind SO in performance, usually by several orders of magnitude. S is, as expected, giving a poor performance unless $t$ is large. Third, the wrapped versions of the pseudo-likelihoods dominate uniformly the unwrapped ones, both having similar performances if the process is highly concentrated. Indeed, the wrapping of SO is key in preventing the spread of probability mass outside $\T^p$ when the Jacobian of the drift has positive eigenvalues and $t$ grows, which raises numerical instabilities (e.g., lower right panel of Figure \ref{fig:klwn2d}). Finally, matching the WN distribution of E and SO by a vM has different effects depending on the method. For E, the results are similar for both E and EvM, except for a bump in small times with high diffusivities. However, SOvM consistently adds a high bias to SO, resulting in significant higher divergences. As a general advice, we recommend to\nopagebreak[4] approximate the tpd of the WN process by WOU, SO and E, in this order. 

\begin{figure}[!h]
	\vspace*{-0.25cm}
	\centering
	\includegraphics[width=.90\textwidth]{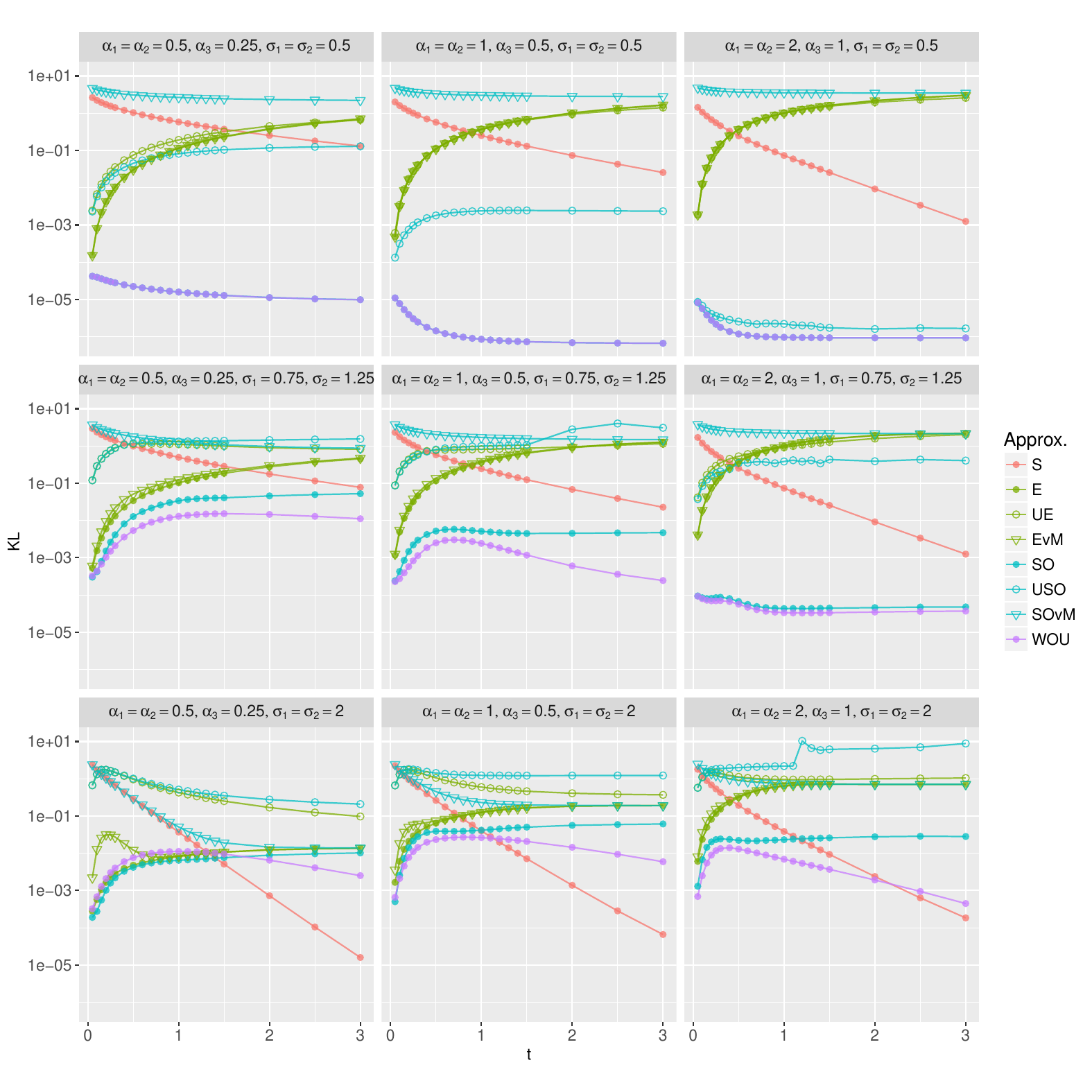}
	\vspace*{-0.5cm}
	\caption{\small $\mathrm{D}^\mathrm{A}_{t,\sigma_0^2}$ curves for $p=2$ on the WN process. Note the vertical log-scale. From top to bottom, rows represent small, moderate, and high diffusivities. The PDE was solved with $M_x=M_y=240$, $M_t=\lceil 1500t\rceil$, and $\sigma_0=0.1$.}
	\label{fig:klwn2d}
\end{figure}

We reproduce the same experiment on the vM process, with results collected in Figures \ref{fig:klvm1d} and \ref{fig:klvm2d}. The highlights are similar except for the following  differences. First, the good properties that WOU has for the WN process do not hold any more, evincing its process-specificity. S is now the only approximation whose accuracy improves over time. Second, SO is systematically above E in performance, yet this difference is reduced as SO is not the true tpd under high-concentration. Third, the vM distribution match does not provide a better approximation to the tpd, despite the sdi being vM. EvM is again close to E and EvM except for small $t$'s where EvM adds a substantial bias for scenarios with moderate and high diffusivities. The same happens for SOvM in $p=1$, whereas for $p=2$ SOvM increases the Kullback--Leibler divergence by several orders of magnitude when compared to SO in the scenarios with high diffusivity. Our general advice is to approximate the tpd by SO and E, in this order.

\begin{figure}[!h]
	\vspace*{-0.25cm}
	\centering
	\includegraphics[width=.90\textwidth]{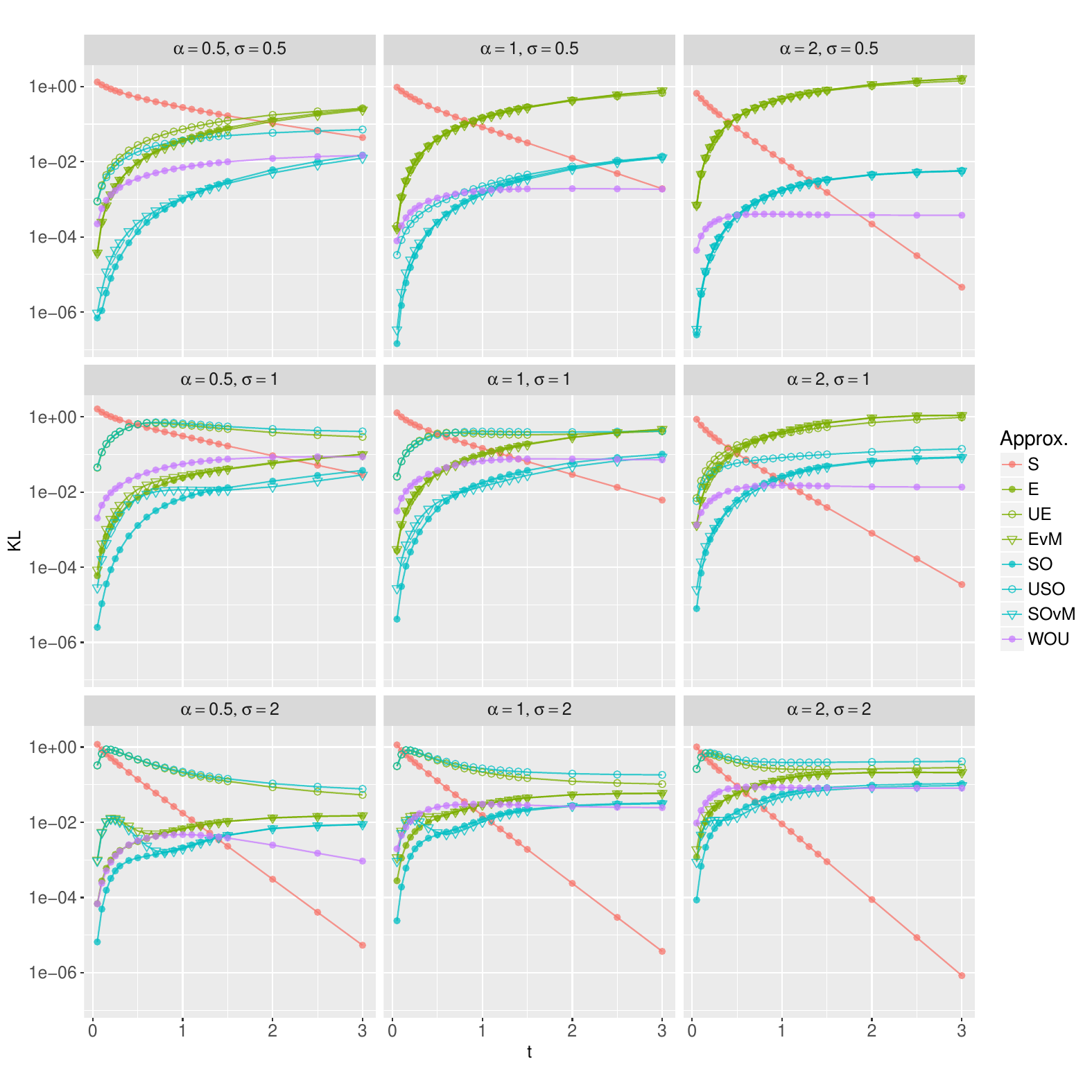}
	\vspace*{-0.5cm}
	\caption{\small $\mathrm{D}^\mathrm{A}_{t,\sigma_0^2}$ curves for $p=1$ on the vM process. The description in Figure \ref{fig:klwn1d} applies.}
	\label{fig:klvm1d}
\end{figure}

%-------------------------------------%
\subsection{Empirical performance of the approximate likelihoods}
%-------------------------------------%

We compare now the efficiency of WOU, SO, and E -- the best performing tpd approximations, according to the weighted Kullback--Leibler divergences -- in estimating the unknown parameters of the diffusion \eqref{eq:param} from a trajectory $\{\bTheta_{\Delta i}\}_{i=0}^N$. In this section, we set $N=250$ and assume that $\sigma(\cdot;\blambda)=\bSigma^\frac{1}{2}$ is known in order to avoid the inherent unidentifiabilities of $\blambda$ when $\Delta$ is large and the tpd converges to the sdi. We explore the behaviour of the estimators for dimensions $p=1,2$, time steps $\Delta=0.05,0.20,0.50,1.00$, and for representative parameter choices of the WN process and of two challenging diffusions. For $p=1$, we also consider the PDE-based approximation to the likelihood. The trajectories are simulated using the E method with time step $0.001$ and then subsampled for given $\Delta$'s. \\

In order to summarize the overall performance of a collection $\{\hat\blambda_j=(\hat\lambda_{j,1},\ldots, \hat\lambda_{j,K}):j=1\ldots,J\}$ of $K$-variate estimators of $\blambda$, we consider a global measure of relative performance. This measure is the componentwise average of Relative Efficiency (RE), where the relative efficiency is measured with respect to the best estimator at a given component in terms of Mean Squared Error (MSE):
\begin{align*}
\mathrm{RE}(\hat\blambda_j):=\frac{1}{K}\sum_{k=1}^K\mathrm{RE}(\hat\lambda_{j,k}|\hat\lambda_{\star,k}),\quad \mathrm{RE}(\hat\lambda_{j,k}|\hat\lambda_{\star,k}):=\frac{\mathrm{MSE}(\hat\lambda_{\star,k})}{\mathrm{MSE}(\hat\lambda_{j,k})},\quad 
\mathrm{MSE}(\hat\lambda_{\star,k}):=\min_{j=1,\ldots,J}\mathrm{MSE}(\hat\lambda_{j,k}).
\end{align*}
Hence, if $\hat\blambda_j$ is the best estimator for all the components of $\blambda$, then $\mathrm{RE}(\hat\blambda_j)=1$. We estimate $\mathrm{RE}(\hat\blambda_j)$ by Monte Carlo with $1000$ replicates, where $\hat\blambda_j$ is obtained by maximizing the approximate likelihood with a common optimization procedure that employs \eqref{eq:smle} as starting values.

\begin{figure}[!h]
	\vspace*{-0.25cm}
	\centering
	\includegraphics[width=.90\textwidth]{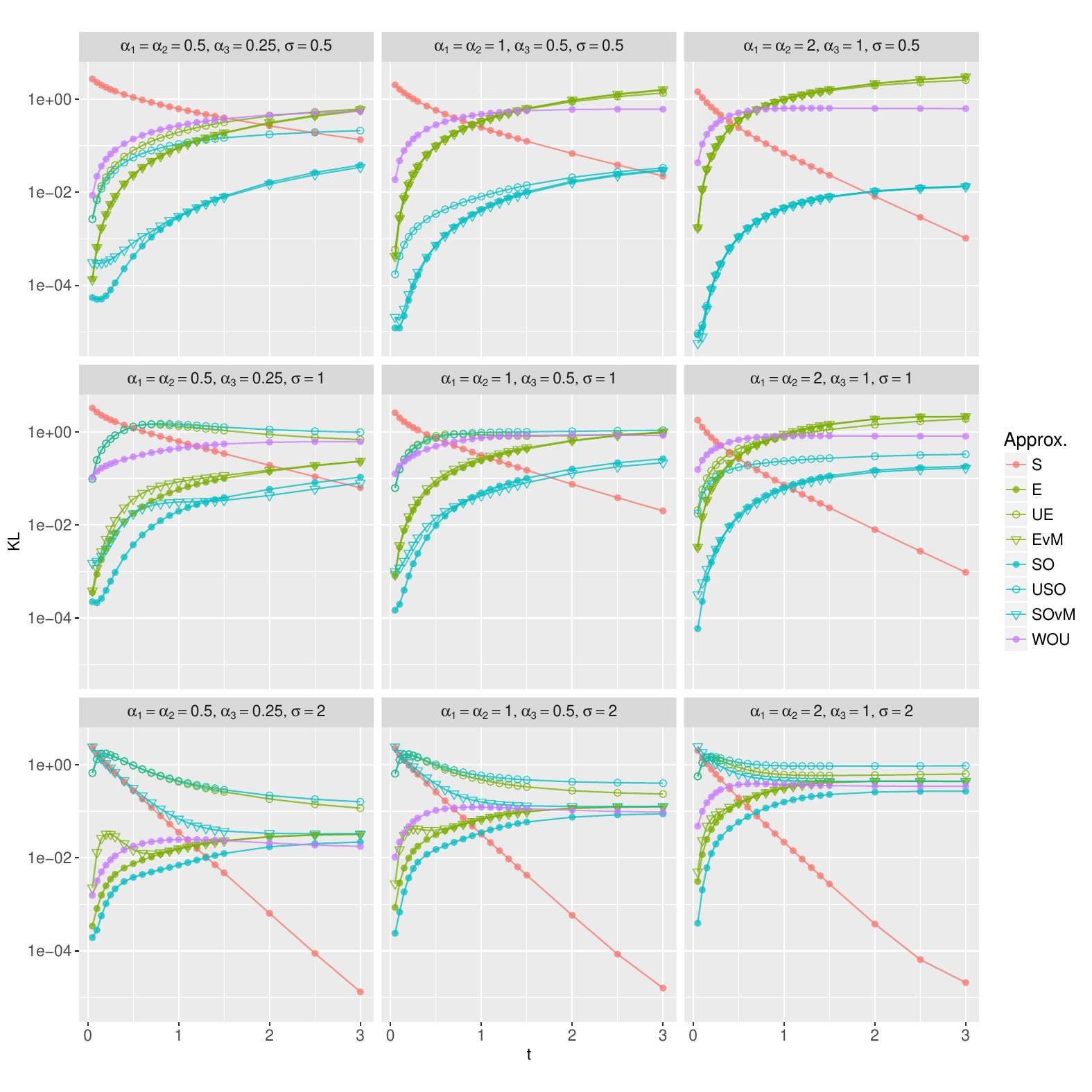}
	\vspace*{-0.5cm}
	\caption{\small $\mathrm{D}^\mathrm{A}_{t,\sigma_0^2}$ curves for $p=2$ on the MvM process. The description in Figure \ref{fig:klwn2d} applies.}
	\label{fig:klvm2d}
\end{figure}

\begin{table}[H]
\centering
\small
\begin{tabular}{c|cccc||cccc}
\toprule\toprule
& \multicolumn{4}{c||}{$\alpha=0.5$,\quad $\sigma=1$} & \multicolumn{4}{c}{$\alpha=1$,\quad $\sigma=1$} \\ \midrule
$\Delta$ &     E     &    SO     &    WOU    &      PDE      &     E     &    SO     &    WOU    &     PDE     \\ \midrule
$0.05$ & \bf  0.9799 &  0.9392 &  0.9608 &  0.7596 & \bf  0.9888 &  0.9229 &  0.9241 &  0.7276\\
$0.20$ & \bf  0.9631 &  0.8554 &  0.8878 &  0.8319 & \bf  0.9937 &  0.8425 &  0.8422 &  0.7852\\
$0.50$ &  0.8941 &  0.7444 &  0.9016 & \bf  0.9340 &  0.6907 &  0.9904 & \bf  1.0000 &  0.9826\\
$1.00$ &  0.5685 &  0.7504 &  0.8978 & \bf  1.0000 &  0.5329 &  0.9763 & \bf  0.9972 &  0.9969\\ \midrule\midrule
& \multicolumn{4}{c||}{$\alpha=0.5$,\quad $\sigma=2$} & \multicolumn{4}{c}{$\alpha=1$,\quad $\sigma=2$} \\ \midrule
$\Delta$ &     E     &    SO     &    WOU    &      PDE      &     E     &    SO     &    WOU    &     PDE     \\ \midrule
$0.05$ &  0.9688 & \bf  0.9972 &  0.9700 &  0.9098 &  0.9392 & \bf  0.9431 &  0.9205 &  0.8760\\
$0.20$ &  0.7586 & \bf  0.9740 &  0.8586 &  0.7805 &  0.8040 &  0.8670 &  0.9319 & \bf  0.9380\\
$0.50$ &  0.6272 &  0.9565 & \bf  1.0000 &  0.8535 &  0.6297 &  0.7321 &  0.8368 & \bf  1.0000\\
$1.00$ &  0.2784 &  0.6904 & \bf  1.0000 &  0.8578 &  0.6090 &  0.8437 &  0.7823 & \bf  0.8964\\ \bottomrule\bottomrule
\end{tabular}
\caption{\small Relative efficiencies for WN diffusion with $p=1$ and $\mu=\tfrac{\pi}{2}$. Boldface highlights the highest relative efficiencies. \label{tab:wn1}}
\end{table}

%-------------------------------------%
\subsubsection{WN process}
%-------------------------------------%\\

Table \ref{tab:wn1} shows the relative efficiencies for E, SO, WOU, and PDE with $p=1$. When averaging across scenarios and discretization times, the global ranking of performance is: WOU ($0.9195$), PDE ($0.8831$), SO ($0.8766$), and E ($0.7642$). On average, E is the best performing method for $\Delta=0.05$, followed closely by SO. However, the relative performance of E severely decays as $\Delta$ increases. A similar pattern is present for SO, although the decay in relative efficiency is less severe, being by a narrow margin the best performing method for $\Delta=0.20$ (above E and WOU with an absolute difference lower than $0.5\%$). PDE is significantly underperforming for $\Delta=0.05,0.20$, which is explained by the bias induced by the initial condition: $\sigma_0=0.1$ was considered as a compromise between tractability ($M_x=500$, $M_t=\lceil100\Delta\rceil$) and accuracy. PDE becomes the best performer on average for $\Delta=0.50,1.00$, where the effects of the initial condition become less important. WOU shows an intermediate profile with an indubitable advantage: on average, its relative efficiency has an absolute difference with respect to the best performing method of less than $2.5\%$. This fact is\nopagebreak[4] what makes it the best method on the global ranking of performance.

\begin{table}[H]
\centering
\small
\begin{tabular}{c|ccc||ccc}
\toprule\toprule
& \multicolumn{3}{c||}{$\alpha=1$,\quad $\sigma=1$} & \multicolumn{3}{c}{$\alpha=2$,\quad $\sigma=1$} \\ \midrule
$\Delta$ &      E      &     SO      &          WOU          &   E    &   SO   &              WOU              \\ \midrule
$0.05$ & \bf  0.9765 &  0.9244 &  0.8999 & \bf  0.9920 &  0.8452 &  0.8460\\
$0.20$ & \bf  0.9985 &  0.8214 &  0.8229 &  0.7234 &  0.9978 & \bf  0.9993\\
$0.50$ &  0.5679 &  0.9868 & \bf  0.9972 &  0.4370 & \bf  1.0000 &  0.9980\\
$1.00$ &  0.4296 &  0.9872 & \bf  0.9998 &  0.3467 & \bf  1.0000 &  0.9970\\ \midrule\midrule
& \multicolumn{3}{c||}{$\alpha=1$,\quad $\sigma=2$} & \multicolumn{3}{c}{$\alpha=2$,\quad $\sigma=2$} \\ \midrule
$\Delta$ &      E      &     SO      &          WOU          &   E    &   SO   &              WOU              \\ \midrule
$0.05$ &  0.9297 & \bf  1.0000 &  0.9422 & \bf  0.9635 &  0.8752 &  0.8793\\
$0.20$ &  0.8249 &  0.9573 & \bf  0.9916 &  0.6017 &  0.7333 & \bf  1.0000\\
$0.50$ &  0.6050 &  0.6607 & \bf  1.0000 &  0.3797 &  0.6406 & \bf  1.0000\\
$1.00$ &  0.5254 &  0.5432 & \bf  1.0000 &  0.2690 &  0.4214 & \bf  1.0000\\ \bottomrule\bottomrule
\end{tabular}
\caption{\small Relative efficiencies for WN diffusion with $p=2$, $\bmu=\lrp{\tfrac{\pi}{2},-\tfrac{\pi}{2}}$, $\alpha_1=\alpha_2=\alpha$, $\alpha_3=\tfrac{\alpha}{2}$, and $\bSigma=\sigma^2\bI$. Boldface highlights the highest relative efficiencies.\label{tab:wn2}}
\end{table}

\pagebreak

Table \ref{tab:wn2} gives the relative efficiencies for E, SO, and WOU in $p=2$. When averaging across scenarios and discretization times, the global ranking of performance is: WOU ($0.9608$), SO ($0.8372$), and E ($0.6607$). Similarly to $p=1$, E is the best performing method for $\Delta=0.05$ and its relative efficiency quickly decays as $\Delta$ increases. SO and WOU perform similarly for low diffusive scenarios ($\sigma=1$), but for $\sigma=2$ WOU significantly outperforms SO for $\Delta=0.20,050,1.00$, a fact explained by the proneness of the tpd to be multimodal in those situations. The competitive performance of WOU for $p=1,2$ under all scenarios and $\Delta$'s, in addition to its affordable computational cost, places it as the preferred estimation method for the WN process.

\begin{table}[H]
\centering
\small
\begin{tabular}{c|llll}
\toprule\toprule
Tpd & \multirow{2}{*}{$t\to0$} & \multirow{2}{*}{$t\in\R^+$} & \multirow{2}{*}{$t\to\infty$} & Comput. \\
approx. &  & &  & expediency \\\midrule
E & \textborn\textborn\textborn\textborn\textborn & \textborn\textborn & \textborn & \textborn\textborn\textborn\textborn\textborn \\
SO & \textborn\textborn\textborn\textborn & \textborn\textborn\textborn & \textborn\textborn\textborn & \textborn\textborn\textborn\\
WOU & \textborn\textborn\textborn\textborn & \textborn\textborn\textborn\textborn & \textborn\textborn\textborn\textborn\textborn & \textborn\textborn\textborn\textborn \\
PDE & \textborn\textborn\textborn & \textborn\textborn\textborn\textborn\textborn & \textborn\textborn\textborn\textborn\textborn & \textborn\\
\bottomrule\bottomrule
\end{tabular}
\caption{\small Comparative of estimation methods for the WN process in $p=1,2$. The number of stars ranges from one to five. The more stars, the better performance in the category. The first three columns give the behaviour of the tpd approximation when $t$ is small, intermediate, and large, respectively. \label{tab:stars}}
\end{table}

%-------------------------------------%
\subsubsection{Other processes}
%-------------------------------------%

The WC diffusion has a remarkably different drift from the WN process (Figure \ref{fig:ld}). As a consequence, the tpd of the WC diffusion quickly becomes highly non-WN (multimodal, ``heavy tails'', peaked), both the opposite defining features of the pseudo-tpds. This affects the relative efficiencies for E, SO, and PDE given in Table \ref{tab:wc}, whose global performance is: PDE ($0.9727$), SO ($0.4587$), and E ($0.4131$). The supremacy of the PDE, except for small drift ($\alpha=0.5$) and $\Delta=0.05$, is evident. Thus, Table \ref{tab:wc} is an illustration of the low efficiency of applying the Euler and Shoji--Ozaki pseudo-likelihoods for highly non-WN processes at arbitrary\nopagebreak[4] $\Delta$'s. \\

\begin{table}[H]
\centering
\small
\begin{tabular}{c|ccc||ccc}
\toprule\toprule
& \multicolumn{3}{c||}{$\alpha=0.5$,\quad $\sigma=1$} & \multicolumn{3}{c}{$\alpha=1$,\quad $\sigma=1$} \\ \midrule
$\Delta$ &     E     &    SO     &      PDE      &     E     &    SO     &     PDE     \\ \midrule
$0.05$ &  0.9277 & \bf  1.0000 &  0.7682 &  0.5715 &  0.5938 & \bf  0.9309\\
$0.20$ &  0.5968 &  0.7315 & \bf  1.0000 &  0.3418 &  0.3524 & \bf  1.0000\\
$0.50$ &  0.3548 &  0.4264 & \bf  1.0000 &  0.2923 &  0.3030 & \bf  1.0000\\
$1.00$ &  0.3068 &  0.3295 & \bf  1.0000 &  0.2865 &  0.2774 & \bf  1.0000\\ \midrule\midrule
& \multicolumn{3}{c||}{$\alpha=0.5$,\quad $\sigma=2$} & \multicolumn{3}{c}{$\alpha=1$,\quad $\sigma=2$} \\ \midrule
$\Delta$ &     E     &    SO     &      PDE      &     E     &    SO     &     PDE     \\ \midrule
$0.05$ & \bf  0.9686 &  0.8947 &  0.8646 &  0.7325 &  0.6734 & \bf  1.0000\\
$0.20$ &  0.8114 &  0.8720 & \bf  1.0000 &  0.0213 &  0.1196 & \bf  1.0000\\
$0.50$ &  0.1867 &  0.3634 & \bf  1.0000 &  0.0258 &  0.0880 & \bf  1.0000\\
$1.00$ &  0.1417 &  0.2396 & \bf  1.0000 &  0.0441 &  0.0750 & \bf  1.0000\\ \bottomrule\bottomrule
\end{tabular}
\caption{\small Relative efficiencies for the WC diffusion ($p=1$) with $\mu=\tfrac{\pi}{2}$. Boldface highlights the highest relative efficiencies.\label{tab:wc}}
\end{table}

Finally, Table \ref{tab:mivm} shows the relative efficiencies of E and SO for a mivM diffusion with antipodal means. In order to avoid spurious maximums, $q$ was estimated by SMLE and then kept fixed when optimizing the approximate likelihood. The global performances are: SO ($0.9655$), and E ($0.8920$). The analysis by $\Delta$'s shows that, as in the WC diffusion, SO is performing better than E except for $\Delta=0.05$. However, inspection of the tpd shows a prevalent multimodality, which points towards a low efficiency of the pseudo-likelihoods when $\Delta$ is not small.

\begin{table}[H]
\centering
\small
\begin{tabular}{c|cc||cc||cc}
\toprule\toprule
& \multicolumn{2}{c||}{$q=0.25$} & \multicolumn{2}{c||}{$q=0.50$} & \multicolumn{2}{c}{$q=0.75$} \\ \midrule
$\Delta$ &      E      &        SO        &      E      &        SO        &      E      &       SO       \\ \midrule
$0.05$ &  0.9282 & \bf  0.9595 & \bf  0.9851 &  0.9620 & \bf  0.9716 &  0.9527\\
$0.20$ &  0.8678 & \bf  0.9901 &  0.8999 & \bf  0.9616 & \bf  0.9517 &  0.9296\\
$0.50$ &  0.8312 & \bf  0.9825 &  0.8223 & \bf  0.9454 &  0.9448 & \bf  0.9640\\
$1.00$ &  0.8867 & \bf  0.9984 &  0.8625 & \bf  0.9742 &  0.7525 & \bf  0.9661\\ \bottomrule\bottomrule
\end{tabular}
\caption{\small Relative efficiencies for the mivM diffusion with $p=2$, $\bM=\lrp{\tfrac{\pi}{2},\tfrac{\pi}{2};-\tfrac{\pi}{2},-\tfrac{\pi}{2}}$,  $\bA=\lrp{\tfrac{3}{4},\tfrac{3}{4};\tfrac{3}{2},\tfrac{3}{2}}$, $\bp=(q,1-q)$, and $\sigma=1$. Boldface highlights the highest relative efficiencies.\label{tab:mivm}}
\end{table}

%-------------------------------------%
\section{Application to molecular dynamics}
\label{sec:app}
%-------------------------------------%

Toroidal data arises from the representation of the backbone of a protein made of $n$ amino acids as a sequence of $n-2$ pairs of dihedral angles $(\phi,\psi)$, thus as a point in $\T^{2(n-2)}$. The dihedral angles capture the rotations around the N--C$_{\alpha}$ and C$_{\alpha}$--C bonds, which are the remaining degrees of freedom of the backbone (if the bond angles and bond lengths are assumed fixed to their ideal values). Molecular dynamics simulations are widely
employed to study the folding and the dynamical properties of proteins,
providing ultra high frequency trajectories of protein structures.
The dihedral angles of the time-varying backbone
result in a trajectory $\{(\phi_{1,i\Delta},\allowbreak\psi_{1,i\Delta},\allowbreak\ldots,\phi_{n-2,i\Delta},\psi_{n-2,i\Delta})\}_{i=0}^{N}$. Diffusive models on the torus are appropriate tools to summarize these trajectories and, once fitted, can be used as computationally affordable emulators of the physical process.\\

We consider data from molecular dynamics simulations of the protein G (Protein Data Bank identifier 1GB1) around its native state. This protein contains $n=56$
amino acids and, due to its relatively small size and availability
of extensive experimental data, is commonly considered in the molecular dynamics literature.
The molecular dynamics simulations were done using the CHARMM36 force field with the
EEF1-SB solvent model \citep{Bottaro2013} during $T=100$ nanoseconds equally discretized in $10000$
time cuts, which afterwards were subsampled to $N=1000$. For
the sake of illustration,
we study two specific trajectories: $\{\psi_{\Delta i}\}_{i=0}^{N}$
of the $9$-th amino acid (Glycine, between Asparagine and Lysine),
and $\{(\phi_{\Delta i},\psi_{\Delta i})\}_{i=0}^{N}$ of the $14$-th amino acid (Glycine, between Lysine and Glutamate). These one- and two-dimensional trajectories exhibit multi- and unimodal patterns that are representative of the general case.\\

The one-dimensional multimodal trajectory was modelled with a diffusion driven by a mixture of two vM distributions, as given in \eqref{eq:mixvm}. The fitting was done with the PDE method with $M_x=500$, $M_t=20$, and $\sigma_0=0.01$. We used SMLE and \eqref{eq:sigma} as starting values, and fixed the mixture proportions to the stationary estimates to avoid spurious minima. The optimization took $115$ seconds in a $1.7$ GHz core for $566$ likelihood evaluations and gave $\hat \ba=(9.06, 5.00)$, $\hat \bmu=(0.23, -2.91)$, $\hat \sigma=1.08$, and $\hat p=0.56$. The first row of Figure \ref{fig:fits} presents a graphical summary of the parametric fit. The first panel shows the observed data and a simulated trajectory from the fitted model, which captures the main patterns of the observed data, except for some outliers.\\

In order to evaluate the goodness-of-fit of the parametric model -- and due to the absence of formal tests directly applicable in this setting, to the best of the authors' knowledge -- we compared graphically the parametric fits of the drift and diffusion coefficient with their nonparametric estimations. To that aim, we considered the following Nadaraya--Watson estimator for the drift
\begin{align}
\hat b_h(\theta):=\sum_{i=0}^{N-1}W_h(\theta,\Theta_{\Delta i})Y_i,\quad W_h(\theta,\Theta_{\Delta i}):=\frac{e^{\cos(\theta-\Theta_{\Delta i})/h^2}}{\sum_{j=0}^{N-1}e^{\cos(\theta-\Theta_{\Delta j})/h^2}},\label{eq:np}
\end{align}
with $Y_i:=\cmod{\Theta_{\Delta(i+1)}-\Theta_{\Delta i}}/\Delta$ and $h$ as the bandwidth parameter. For the diffusion coefficient, we set $Y_i:=\lrp{\cmod{\Theta_{\Delta(i+1)}-\Theta_{\Delta i}}}^2/\Delta$ and then took the square root in the estimate. To remove the smoothing bias of \eqref{eq:np}, we smoothed the parametric estimate by considering $Y_i=b(\Theta_{i\Delta};\hat\blambda)$ in \eqref{eq:np}, hence equating both biases under the correct specification of the model. The second panel in first row of Figure \ref{fig:fits} compares the nonparametric and smoothed parametric estimates of the drift. Both drifts are shadowed according to a kernel density estimate that emphasizes the regions were the data is present. For those regions, there is a close match between both estimates. The third panel shows a similar analysis for the diffusion coefficient, whose nonparametric estimate exhibits mild departures from $\hat\sigma$ in the regions with high density. 

\begin{figure}[H]
	\centering
	\includegraphics[width=0.33\textwidth]{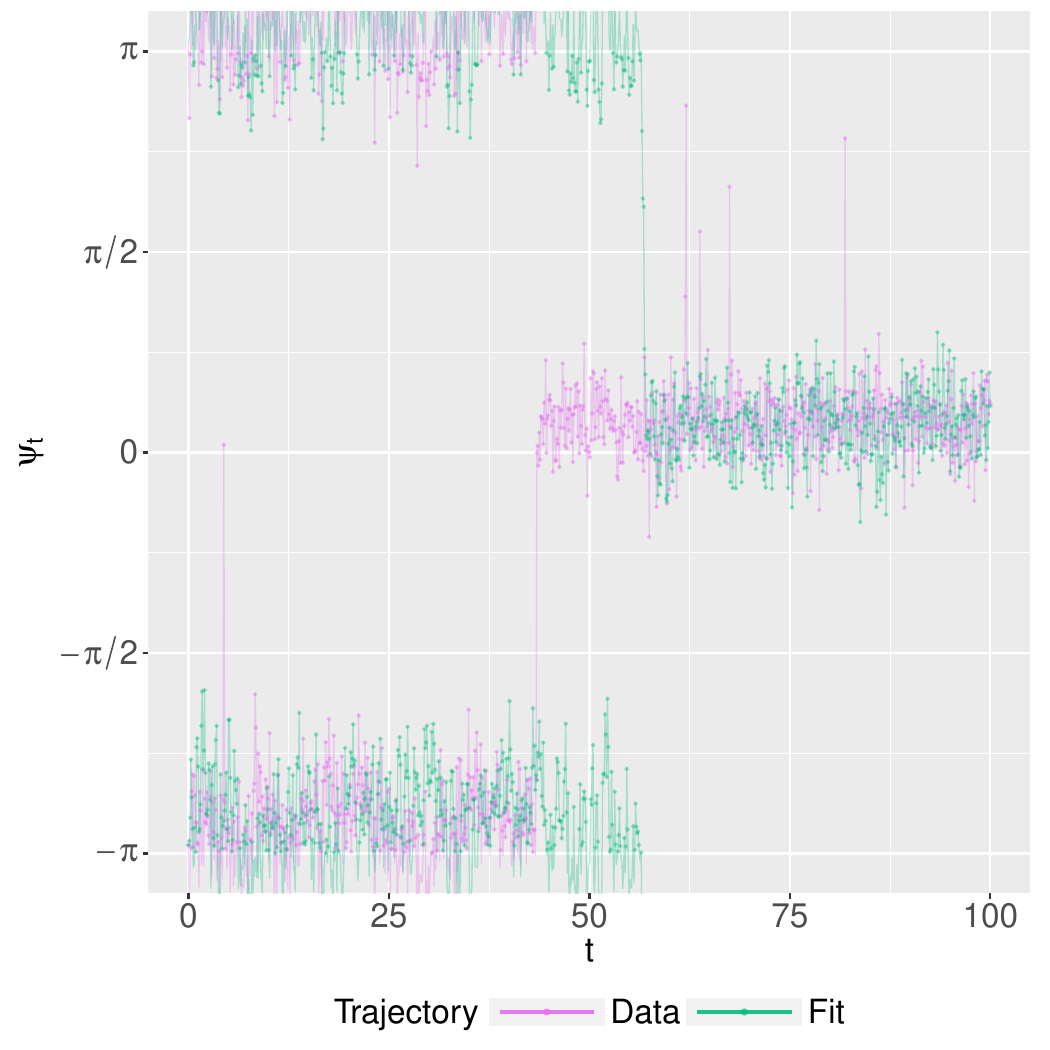}\includegraphics[width=0.33\textwidth]{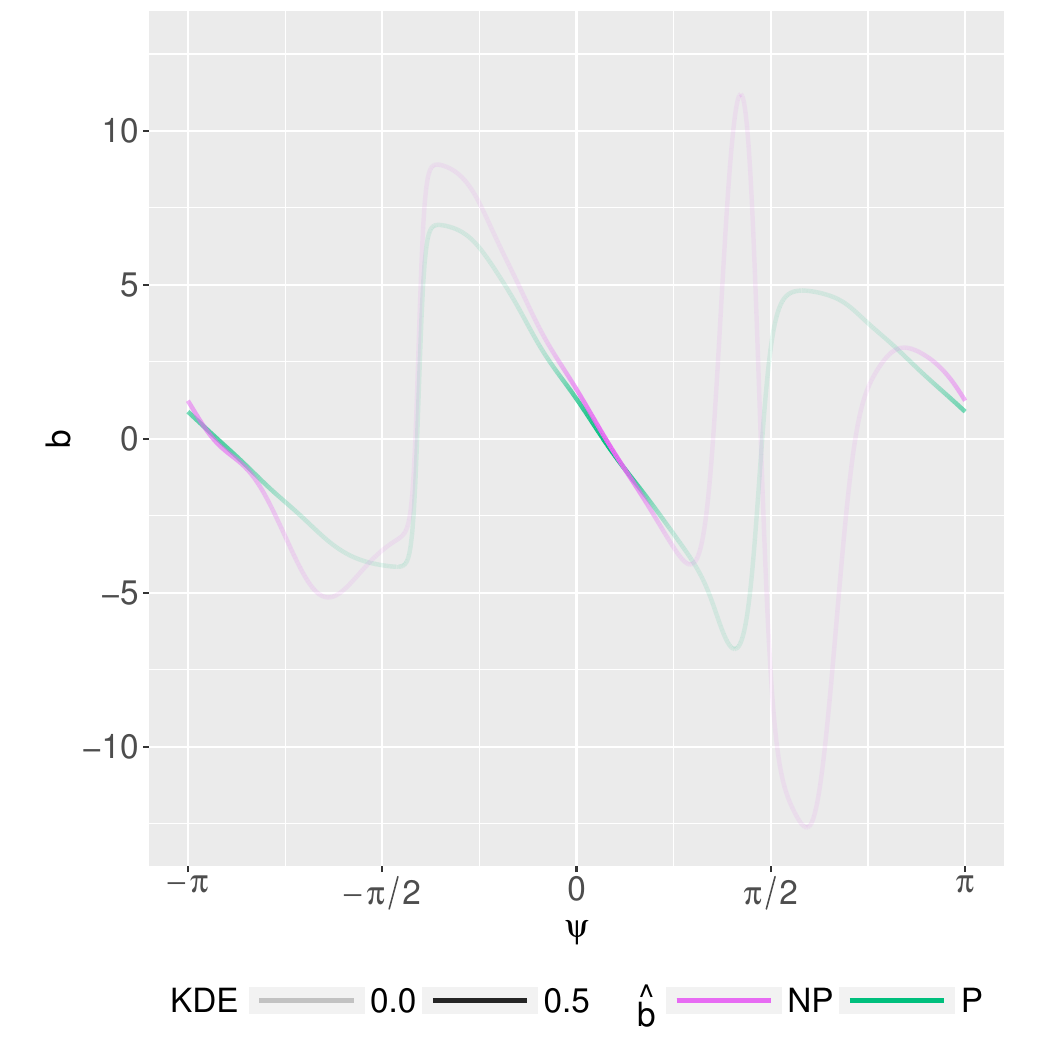}\includegraphics[width=0.33\textwidth]{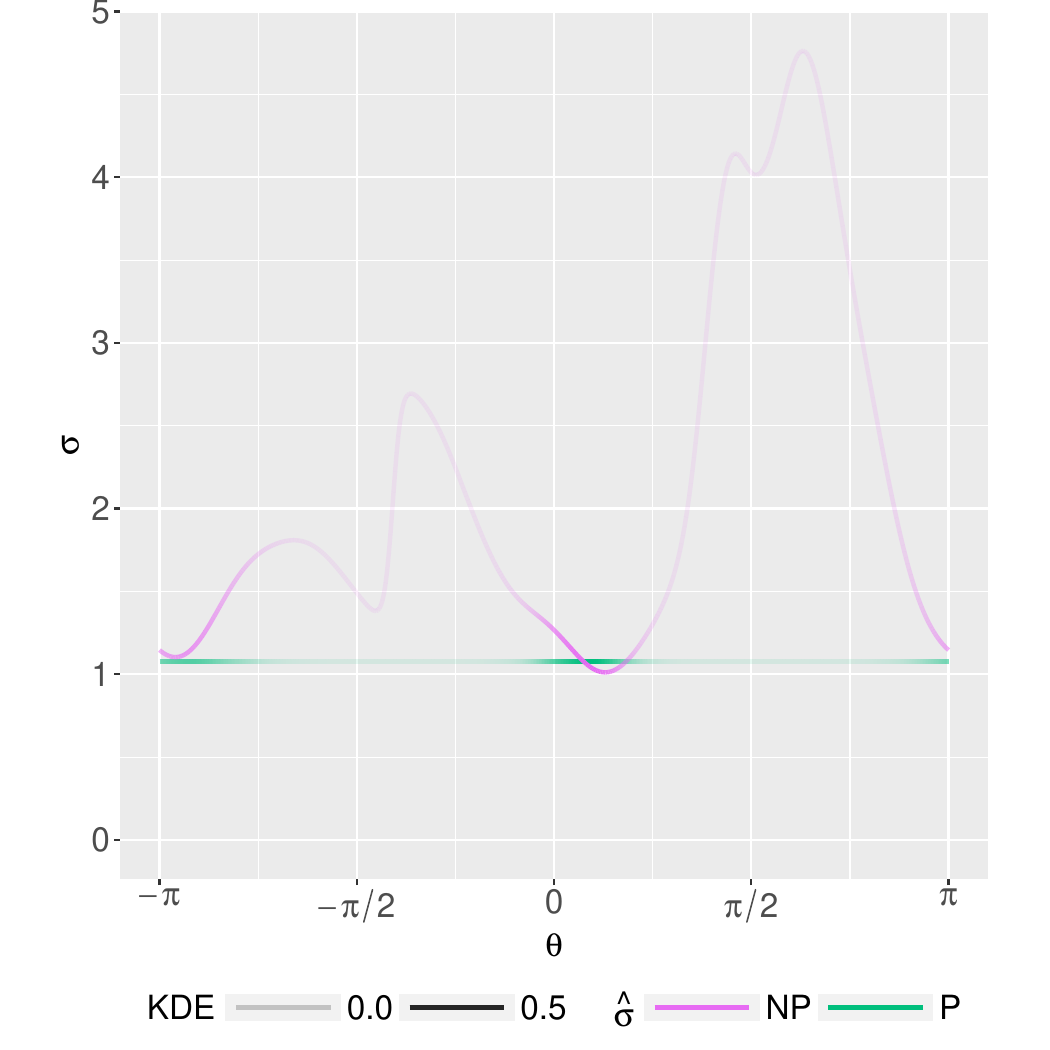}\\
	\includegraphics[width=0.33\textwidth]{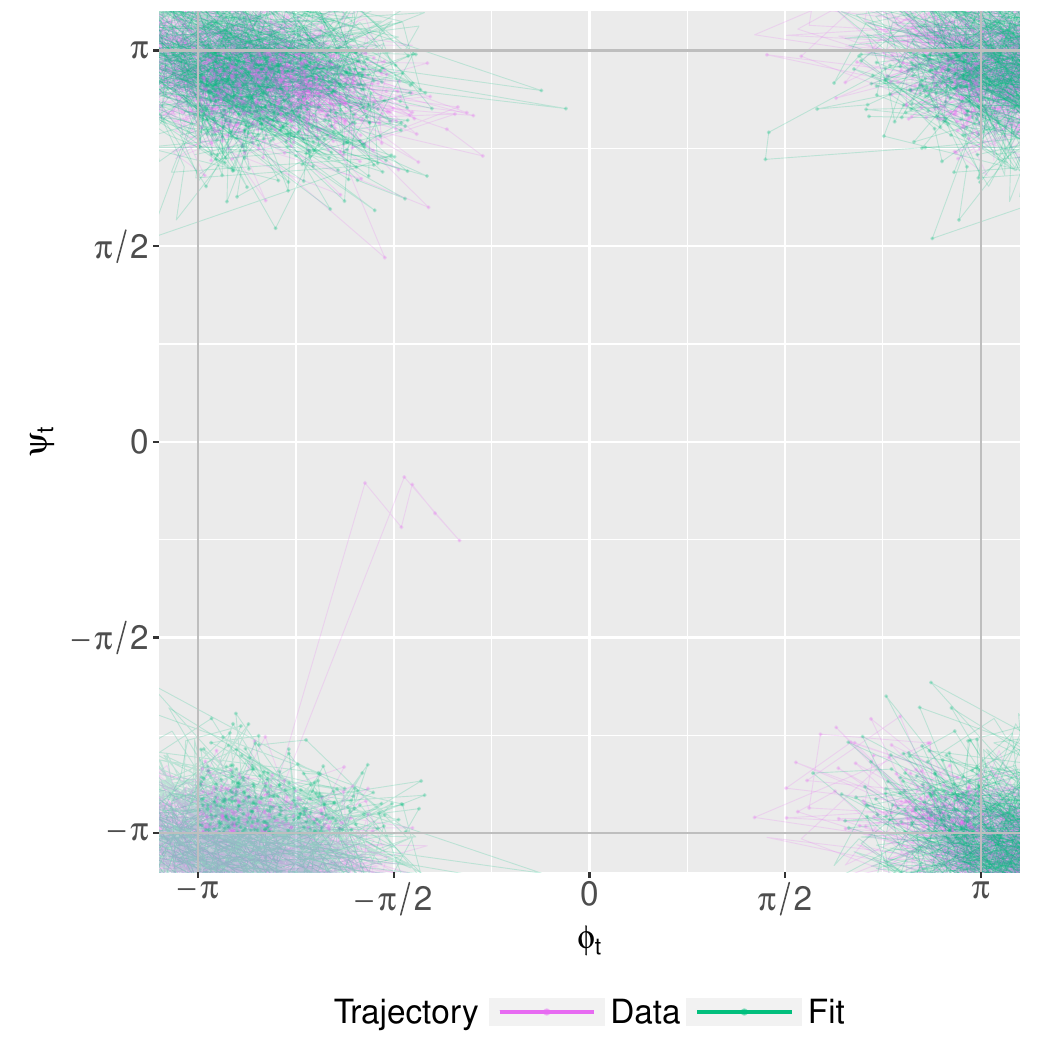}\includegraphics[width=0.33\textwidth]{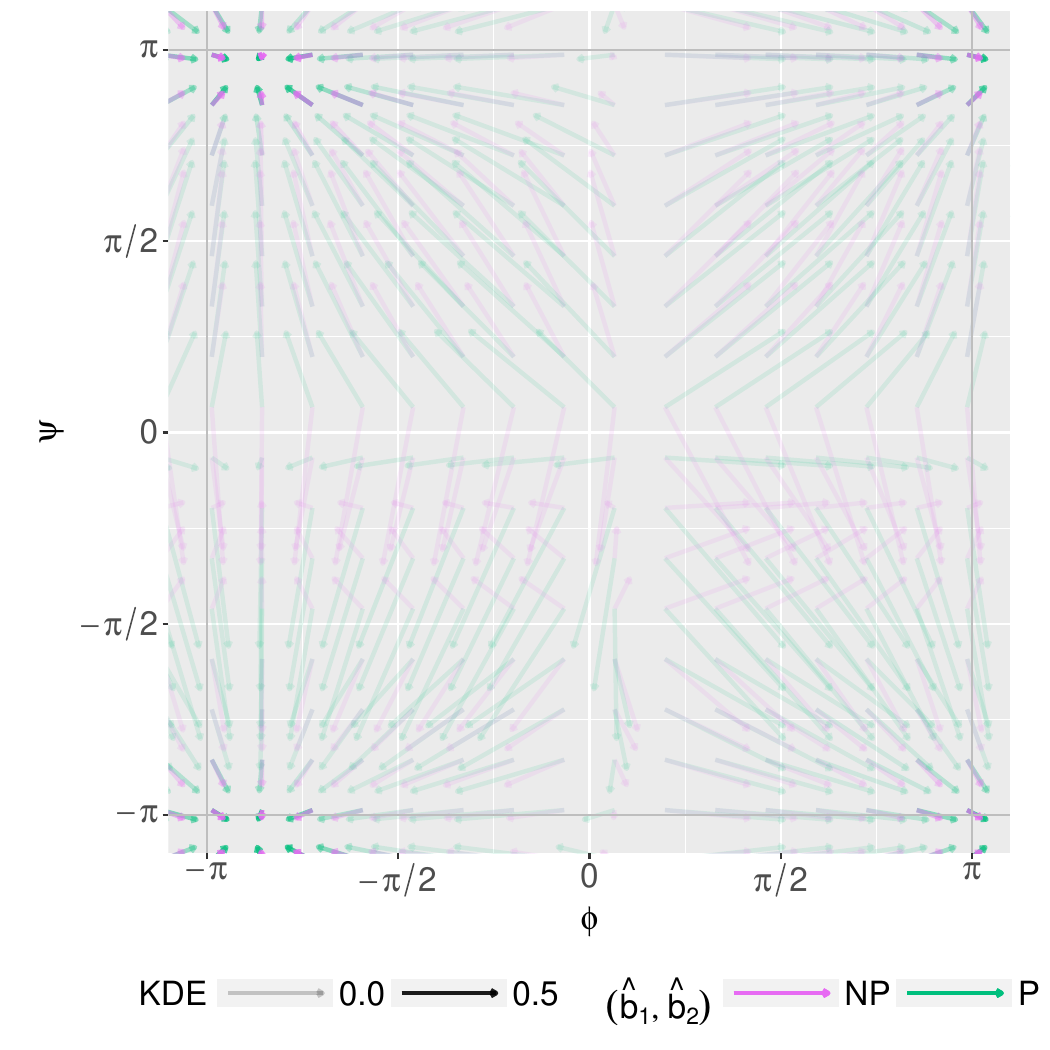}\includegraphics[width=0.33\textwidth]{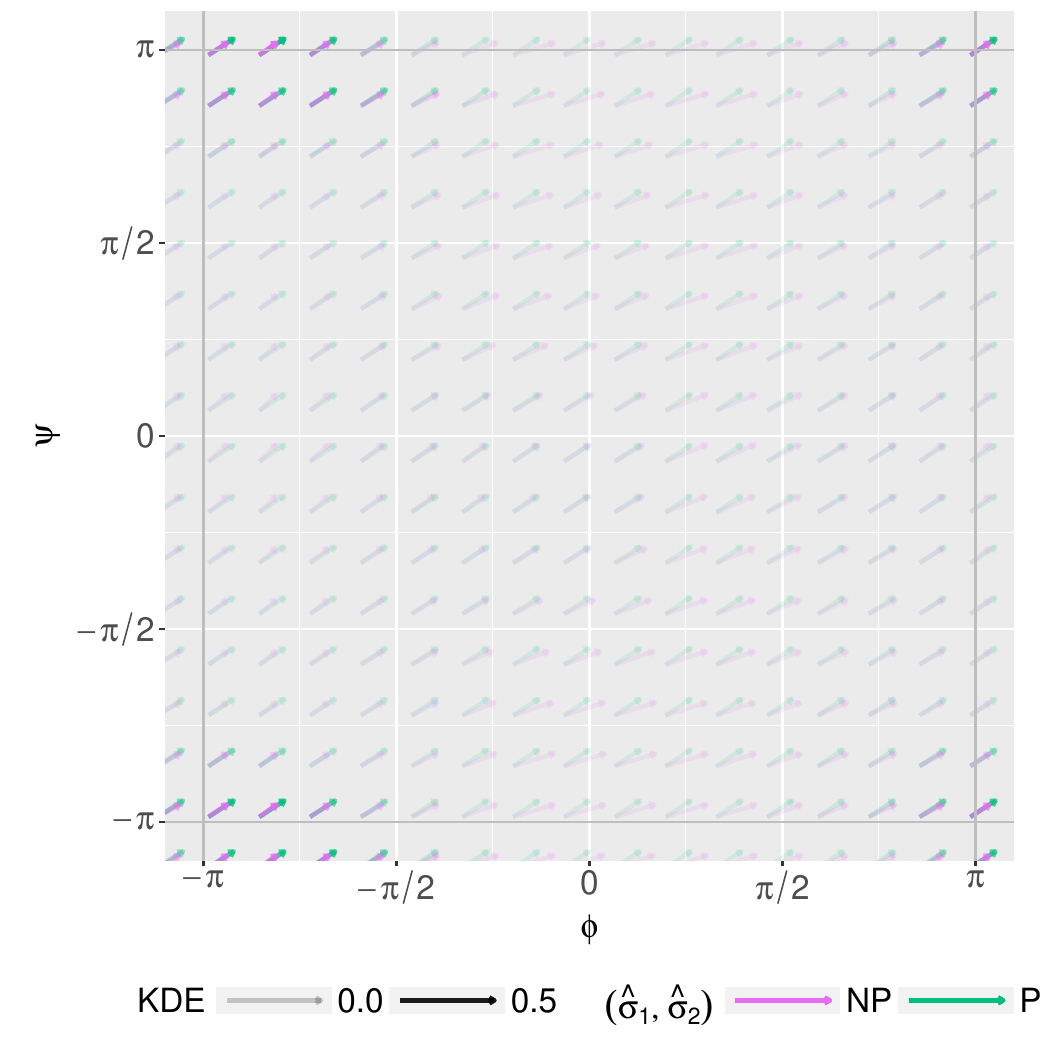}
	\vspace*{-0.3cm}
	\caption{\small
		Summary plots for the fits of $\{\psi_{\Delta i}\}_{i=0}^N$ (first row; $9$-th amino acid) and $\{(\phi_{\Delta i},\psi_{\Delta i})\}_{i=0}^N$ (second row; $14$-th amino acid). From left to right, by columns: observed trajectory and a sample from the fitted model; NonParametric (NP) and Parametric (P) drift estimates; NP and P diffusion estimates. Shading is done according to a KDE of the observed data.\label{fig:fits}}
\end{figure}

For modelling the two-dimensional and unimodal trajectory we employed a bivariate WN diffusion with unconstrained $\bSigma$. The fitting was done with the WOU approximation using SMLE and \eqref{eq:sigma} for starting values. The optimization took $14$ seconds for $2739$ approximate likelihood evaluations. The first panel in the second row of Figure \ref{fig:fits} shows the correct match between the simulated and the observed trajectories, again except for some outliers from the latter. The next panel shows the comparison between the vector fields for the smoothed parametric and nonparametric drifts. They show a strong agreement on the drift structure at regions with presence of data, both in magnitude and direction. The parametric vector field $(\sigma_1(\phi,\psi;\hat\blambda),\sigma_2(\phi,\psi;\hat\blambda))$ and the nonparametric $(\hat\sigma_{1,h_1}(\phi,\psi),\hat\sigma_{2,h_2}(\phi,\psi))$ have a proper match for the regions with data, the latter being constant in most of $\T^p$. The nonparametric estimates were constructed by considering product kernels on the covariates. All the bandwidths were automatically selected by cross-validation.

%-------------------------------------%
\section{Conclusions}
\label{sec:conc}
%-------------------------------------%

We introduced ergodic diffusions on the torus as the natural processes with stationary distributions equal to well-known toroidal distributions. The WN process, with an available analytical approximation to its tpd, is shown to be the most tractable OU-like toroidal process among the different proposals. This approximation outperforms the wrapped Euler and Shoji--Ozaki pseudo-likelihoods, and shows an affordable computational cost for one and two dimensions. In addition, we provide numerical solutions of the one- and two-dimensional Fokker--Planck PDEs for approximating the true tpd, which serve as benchmarks of the accuracy of the approximating tpds. A thorough simulation study explored the performance of the approximate likelihoods under different scenarios. Finally, a data application illustrated the usefulness of the new diffusive models for modelling molecular dynamics simulations. \\

We summarize some important practical conclusions. For estimating the WN process, we recommend to use WOU as a first option for a fast and accurate approximation in dimensions $p=1,2$. For a general process, we advise to employ PDE with $p=1$ if accuracy is a priority, and SO in case speed is. For $p=2$, SO is preferred to E, but both are prone to underperform severely for highly non-WN tpds, which can be visualized using the PDE solution. \\

The development of a general and computationally fast method
for approximating an arbitrary tpd, that is able to cope with multimodality,
remains an open challenge. A promising avenue is methods based on
simulation, which have been successful for Euclidean diffusions; see e.g.
\cite{Beskos2006}, \cite{Papaspiliopoulos2012}, \cite{Sermaidis2012}, \cite{Bladt2016},
and references in these papers. The simplest algorithm by \cite{Beskos2006a} is well suited for exact simulation of the transient diffusion (\textit{i.e.}, before wrapping) because of the periodicity
of the coefficients, and the method in \cite{Sermaidis2012} is applicable
to Langevin diffusions. It is therefore likely that the exact
simulation methods can be adapted to toroidal Langevin diffusions
by finding ways to deal with the wrapping when simulating diffusion
bridges. It is also of interest to study whether the coupling methods
underlying the diffusions bridge simulation technique in
\cite{Bladt2016} can be adapted to the torus setting.
Another interesting approach would be to include the winding number
for each observation as a latent variable and apply methods like the
EM algorithm or the Gibbs sampler for likelihood inference.

%-------------------------------------%
\section*{Software}
\label{sec:soft}
%-------------------------------------%

The software \texttt{sdetorus}, available at \url{https://github.com/egarpor/sdetorus}, contains the implementations of the methods described in the paper and the files required for reproducing all the empirical analyses.

%-------------------------------------%
\section*{Acknowledgements}
%-------------------------------------%

This work is part of the Dynamical Systems Interdisciplinary Network, University of Copenhagen. It was funded by the University of Copenhagen 2016 Excellence Programme for Interdisciplinary Research (UCPH2016-DSIN) and by project MTM2016-76969-P from the Spanish Ministry of Economy, Industry and Competitiveness, and European Regional Development Fund (ERDF). We acknowledge the insightful discussions with John Kent, Jotun Hein, and Michael Golden that led to the key motivation for the manuscript. We are grateful to Sandro Bottaro for the providing the molecular dynamics data used in the illustration. We acknowledge the valuable comments and remarks provided by two anonymous referees and an Associate Editor, which significantly improved the manuscript.

\end{document}